\documentclass{IEEEtran}

\usepackage{graphicx,amsthm,amsmath,amsfonts,latexsym,amssymb}
\usepackage{color,array,verbatim,algpseudocode,bm}
\definecolor{menucolor}{rgb}{0.1,0.52,0.47}
\definecolor{urlcolor}{rgb}{0.85,0.37,0.01}
\definecolor{runcolor}{rgb}{0.46,0.44,0.701}
\definecolor{filecolor}{rgb}{0.2,0.5,0.01}
\definecolor{linkcolor}{rgb}{0.12,0.47,0.70}
\definecolor{citecolor}{rgb}{0.55,0.36,0.01}
\definecolor{anchorcolor}{rgb}{0.4,0.4,0.4}
\usepackage[colorlinks=true, urlcolor=urlcolor, linkcolor=linkcolor,citecolor=citecolor,filecolor=filecolor, anchorcolor=anchorcolor,menucolor=menucolor]{hyperref}
\usepackage{nameref}
\usepackage{zref-xr}
\usepackage{cite}
\zxrsetup{
  tozreflabel=false,
  toltxlabel=true,  
}
\usepackage[shortlabels]{enumitem}
\ifCLASSOPTIONcompsoc
  \usepackage[caption=false,font=normalsize,labelfont=sf,textfont=sf]{subfig}
\else
\usepackage[caption=false,font=footnotesize]{subfig}
\fi
\newtheorem{theorem}{Theorem}

\newtheorem{result}[theorem]{Result}

\usepackage{colortbl}
\usepackage{graphicx}
\usepackage{subfig}
\usepackage{verbatim}
\usepackage{amssymb}
\usepackage{amsmath}
\usepackage{amsthm}
\usepackage{multirow}
\usepackage{rotating}
\usepackage{wrapfig}
\usepackage{placeins}
\usepackage{afterpage}
\definecolor{menucolor}{rgb}{0.1,0.52,0.47}
\definecolor{anchorcolor}{rgb}{0.85,0.37,0.01}
\definecolor{runcolor}{rgb}{0.46,0.44,0.701}
\definecolor{linkcolor}{rgb}{0.3,0.55,0.01}
\definecolor{urlcolor}{rgb}{0.12,0.47,0.70}
\definecolor{citecolor}{rgb}{0.55,0.36,0.01}
\definecolor{filecolor}{rgb}{0.4,0.4,0.4}

\newcommand{\abs}[1]{|#1|}

\newcommand{\balpha}{\boldsymbol{\alpha}}

\newcommand{\bbeta}{\boldsymbol{\beta}}

\newcommand{\bPi}{\boldsymbol{\Pi}}
\newcommand{\bpi}{\boldsymbol{\pi}}

\newcommand{\bGamma}{\boldsymbol{\Gamma}}

\newcommand{\bSigma}{\boldsymbol{\Sigma}}

\newcommand{\bmu}{\boldsymbol{\mu}}

\newcommand{\bTheta}{\boldsymbol{\Theta}}

\newcommand{\bOmega}{\boldsymbol{\Omega}}

\newcommand{\bp}{\boldsymbol{p}}

\newcommand{\bW}{\boldsymbol{W}}

\newcommand{\bv}{\boldsymbol{v}}

\newcommand{\bx}{\boldsymbol{x}}
\newcommand{\bX}{\boldsymbol{X}}

\newcommand{\bone}{\boldsymbol{1}}

\newcommand{\mJ}{\mathcal J}

\newcommand{\mR}{\mathcal R}

\newcommand{\argmax}{\operatornamewithlimits{argmax}}

\newcommand{\ben}{\begin{enumerate}}
\newcommand{\een}{\end{enumerate}}

\newcommand{\E}{\mathbb{E}}
\newcommand{\mB}{\mathcal B}
\newcommand{\mE}{\mathcal E}
\newcommand{\pkg}[1]{{\fontseries{b}\selectfont #1}}
\makeatletter
\newcommand\code{\bgroup\@makeother\_\@makeother\~\@makeother\$\@codex}
\def\@codex#1{{\normalfont\ttfamily\hyphenchar\font=-1 #1}\egroup}
\makeatother
\let\proglang=\textsf

\newcommand{\citep}{\cite}
\newcommand{\citet}{\cite}
\begin{document}
\title{A practical model-based segmentation approach for improved activation detection in   single-subject functional Magnetic Resonance Imaging studies}
\author{Wei-Chen~Chen~and~Ranjan~Maitra
}




\maketitle
\begin{abstract}
       Functional Magnetic Resonance Imaging (fMRI) maps   cerebral
  activation in response to stimuli but this activation is often
  difficult to detect, especially in low-signal contexts and
  single-subject studies.
  Accurate activation detection can be guided by the fact that 
  very few voxels   are, in reality, truly activated and that
  these voxels are spatially localized, but it is challenging to
  incorporate both these   facts.
  We address these twin challenges to single-subject and low-signal
  fMRI by developing a computationally feasible and methodologically
  sound model-based approach, implemented in the R package
  \pkg{MixfMRI}, that  bounds 
  the   {\em a priori}   expected   proportion of activated voxels while also 
  incorporating spatial context.
  An added benefit of our methodology is the ability to distinguish
  voxels and regions having different intensities of activation.
  Our suggested approach is evaluated in realistic two- and three-dimensional
  simulation experiments as well as on multiple real-world datasets. Finally, the
  value of our suggested approach in low-signal and   single-subject
  fMRI 
  studies is illustrated on a sports imagination  experiment that is often used to  detect awareness and improve  treatment in  patients in
  persistent  vegetative state (PVS). Our ability to reliably distinguish activation in
  this experiment potentially opens the door to the adoption of fMRI as a 
  clinical tool for the improved treatment and
  therapy of PVS survivors and other patients.

   \end{abstract}

\maketitle

\section{Introduction}
\label{introduction}
Functional Magnetic Resonance Imaging (fMRI) is an imaging tool 
for understanding the spatial characteristics of human cognitive and motor function~\citep{belliveauetal91, kwongetal92,
  bandettinietal93, howsemanandbowtell98,lindquist08,lazar08}. It is
predicated on the significant discovery over three decades ago
\citep{ogawaetal90a} that the firing of neurons in 
response to the application of a stimulus or the performance of a
task is accompanied by changes in the blood oxygen levels in
neighboring vessels, yielding the so-called Blood Oxygen Level
Dependent (BOLD) contrast and that this BOLD effect is also present in
human brains~\citep{ogawaetal90b}. The BOLD effect causes a shift
in the MR signal and is used in fMRI as a surrogate for neural
activity.  Images of BOLD measurements at each voxel -- or volume
element -- are acquired at multiple times during or without the task- or
stimulus-related activity and, before preprocessing to address the
effects of motion and other artifacts, 
form the raw data from an fMRI experiment. 

An important outcome of many fMRI experiments is the
construction of voxel-wise maps to identify regions of neural activation. Such
activation maps are often obtained  by fitting a general linear 
model~\citep{fristonetal95} to the afore-mentioned raw data, that is, the BOLD time series
observations at each voxel. Statistical Parametric Mapping 
\citep{fristonetal90} then reduces the data at each voxel to a 
test statistic that 
summarizes the association between the BOLD  response and the stimulus
time-course~\citep{bandettinietal93}. The test statistics
(or their $p$-values) are  thresholded to determine
significantly activated voxels~\citep{worsleyetal96, genoveseetal02,loganandrowe04,monti11}. Many factors
~\citep{biswaletal96,hajnaletal94} challenge detection
accuracy,
leading to much variation in identified activation from one study to
another even when the same subject is scanned under the same
paradigm~\citep{maitraetal02,gullapallietal05,maitra09b}.
Improving
detection accuracy is important for 
clinical adoption of fMRI to understand and evaluate individual cognitive
functions and to aid diagnosis by identifying pathologies. 

There are many thresholding approaches 
to control  the False Discovery Rate
(FDR)~\citep{genoveseetal02,nicholsandhayasaka03,benjaminiandheller07,benjaminiandheller08}, with refinements~\citep{benjaminiandheller07} for weighted FDR
to include spatial context and  determine activation of 
clusters of voxels. Other approaches~\citep{fristonetal91,worsleyetal92,worsley94,worsleyandtaylor05} use Random Field Theory 
and specified correlation structure on test statistics to calculate
the threshold that produces an expected Euler
characteristic~\citep{adler81} equal to 
the required $p$-value~\citep{worsley94,worsleyetal96}. Cluster-wise
thresholding is perhaps the most popular approach and involves drawing 
clusters of a minimum number of connected voxels having 
$p$-values below a specified 
threshold~\citep{fristonetal94,hayasakaandnichols03}.  
The method requires specification of neighborhood order, thresholding
$p$-value and minimum cluster size. There has been
 little study on how
 neighborhood order impacts activation  detection accuracy, but many
 researchers use a second-order neighborhood. 
 Earlier $p$-value 
threshold recommendations~\citep{fromanetal95,lazar08} of between
0.02 and 0.03 were deemed too liberal in~\citet{wooetal14} who suggested 
a more conservative default of 0.001. Threshold-free cluster
enhancement (TFCE) methods~\citep{smithandnichols09,helleretal06,spisaketal19}
obviate the need for 
threshold choice but can have low spatial specificity for larger regions
or need precise {\em a priori} information~\citep{wooetal14}.
FDR in cluster-wise thresholding is
controlled by choosing a minimum cluster size based on simulations
under a null-activation model, but recent empirical studies have indicated 
cluster-wise thresholding implementations of  popular fMRI software to
have inflated false positive rates, both in single
subject~\citep{eklundetal15} and group~\citep{eklundetal16}
studies, though corrections and
re-evaluation~\citep{eklundetal16,kessleretal17,coxetal17} have since
indicated that the conclusion may have been unnecessarily alarmist. 

The \citet{brain2025} highlighted the  need for holistic approaches
that account for the spatial, 
temporal and behavioral properties of the experiment and to improve SNR
to make reliable inference at the single-subject level while also
moving the field to identify individual pathologies and
help improve patient care in a clinical setting. In this context,
\citet{brownandbehrmann17} called for the linking of statistical
methodology and the fMRI research paradigm to develop more accurate
analysis methods, ensuring reliability and reproducibility. 

One currently unexploited fact of fMRI activation studies is
that only a very small~\citep{helleretal06,lazar08,almodovarandmaitra19} proportion (about 1-3\%) of in-brain voxels are
{\em a priori} expected to be truly activated. 
So, in this paper, we incorporate this information along with spatial
context to improve detection accuracy even at the single-subject
level in a computationally practical framework with the following objectives:
\begin{itemize}
\item Constrain {\em a priori}, the expected proportion
  of activated voxels. Importantly, this means {\em we do not
    strictly constrain the     actual proportion of activated voxels in a
    particular study}, but rather, {\em we constrain the expected proportion of
    active voxels} to reflect our prior beliefs. 
\item Promote spatial contiguity and context in the detected
  activation so that activated voxels occur in clusters. 
\end{itemize}
It is common to address the latter goal through the use of smoothing
or through other devices such as a spatial mixture
model~\citep{hartvigandjensen00}, Markov Random Field priors such as
the Ising  
model~\citep{genovese00,smithandfahrmeir07}, empirical Bayesian
methods~\citep{fristonetal02a,fristonetal02b} or Bayesian hierarchical modeling~\citep{bowmanetal08,ahnetal11}: however, these approaches
have not been successfully coupled with our first objective that places a limit
on the {\em a priori} expected 
proportion of activated voxels. Our approach postulates a mixture
model for the distribution of the $p$-value of the test statistic at
each voxel -- the first mixture 
component represents the inactive voxels and has mixing 
proportion set to be no less than the {\em a priori} expected
proportion of inactive voxels. Spatial contiguity in the detected
activation is encouraged through a practical penalty term added to the loglikelihood function. Our showcase
application, introduced in Section~\ref{sec:data}, lays out the challenges in accurately detecting
activation in a  sports imagination experiment, illustrating both
the need and value of our methodology because of its implications in
moving fMRI to improve treatment and therapy of, for example,
persistent vegetative state (PVS) patients.



The remainder of this paper is organized as follows. We provide
practical methods for accurate activation detection in single-subject
fMRI and discuss its implementation in 
Section~\ref{mainsec:stat}. We next comprehensively evaluate performance in
simulation experiments in Section~\ref{sec:simulations}
 for a variety of levels and problem
 difficulty,  in 2D, and different noise levels in 3D. We also
 evaluate performance on many standard datasets in the literature. Section~\ref{sec:applications} analyzes the 
sports imagination dataset introduced in Section~\ref{sec:data}. The paper
 concludes with some discussion in Section~\ref{sec:discussion}.
We also have appendices providing additional details on the preprocessing of our
 dataset, statistical methodology, experimental illustrations,
 performance evaluations, and data analysis.

\section{Improving treatment of patients in persistent vegetative state via sports imagination experiments}
\label{sec:data}
\citet{taskforce94} reported that some PVS patients  are known to retain some cognitive
functions~\citep{schiffetal02}. 
Indeed, the supplementary motor area (SMA) was found to be
significantly affected in the fMRI study~\citep{owenetal06} of a
25-year-old female Traumatic Brain Injury (TBI) survivor in 
PVS while she was 
verbally instructed to alternately rest and imagine playing
tennis. Similar activation was   
observed in 12 healthy volunteers, so it has been
surmised~\citep{owenetal06} and demonstrated~\citep{montietal10} that
fMRI can be used to diagnose conscious awareness and to
communicate with PVS patients. A more elaborate
study~\citep{bardinetal11} found  consistent activation (beyond the
SMA, into the parietal cortex and other regions) in 14 normal
subjects, but murky 
consistency  
in seven TBI patients.  While healthy subjects necessarily have
cognitive function that is mostly intact and  regionally consistent in standardized ({\em e.g.}, Talaraich) space, 
TBI patients with injuries impacting brain structure and
function may be missing some standardized regions, or show
inconsistent disruptions across TBI-afflicted brains. 
It is therefore important that activation detection be reliable and
accurate at the individual subject level, and achieving these goals
is the rationale behind our proposed methodology.

We illustrate the challenges to accurate detection in
  single-subject fMRI experiments, by  considering a healthy 40-year-old
female  volunteer~\citep{bardinetal11} who was instructed to
alternately rest or imagine playing tennis, in blocks of 30 seconds
each, and with the entire 210-second study starting and ending
with a resting block. \citet{bardinetal11} do not mention analysis of
this dataset further, but the raw data were publicly
released by~\citet{tabelowandpolzehl11} to demonstrate features of
their \pkg{fmri}  package in {\sf R}~\citep{Rcore}.
We refer to Appendix~\ref{sec:data.supp} for details on data
collection, preprocessing and analysis but note here that several
methods, for example, after controlling for FDR at
$q=0.05$~\citep{genoveseetal02,benjaminiandyekutieli01}, permutation-based testing~\citep{winkleretal14},
thresholding by the value that produces an expected
Euler characteristic~\citep{adler81,worsley94} equal to the desired p-value (taken to be 0.05
in this article), highlighting the challenges faced by global thresholding
methods in accurately and reliably constructing activation maps in
low-signal single-subject experiments. Such was also the result using
spatial mixture modeling~\citep{hartvigandjensen00}. We also employed
cluster-wise thresholding methods to detect activation, with the
minimum number of contiguous voxels in each activated region
determined using the \proglang{3dClustSim} function of the Analysis of
Functional Neuroimages (AFNI) software~\citep{cox96,coxandhyde97,cox12}
 at a threshold of 0.001
(following the recommendations of \citep{wooetal14}) and using first-, second- and third-order
neighborhoods. These three orders of
neighborhood correspond to the definition of neighboring
voxels accordingly as when their (1) faces,
(2) faces or edges, and (3) faces, edges or corners touch. All 
 \begin{figure}[h]
   \subfloat[]{\includegraphics[width=\linewidth]{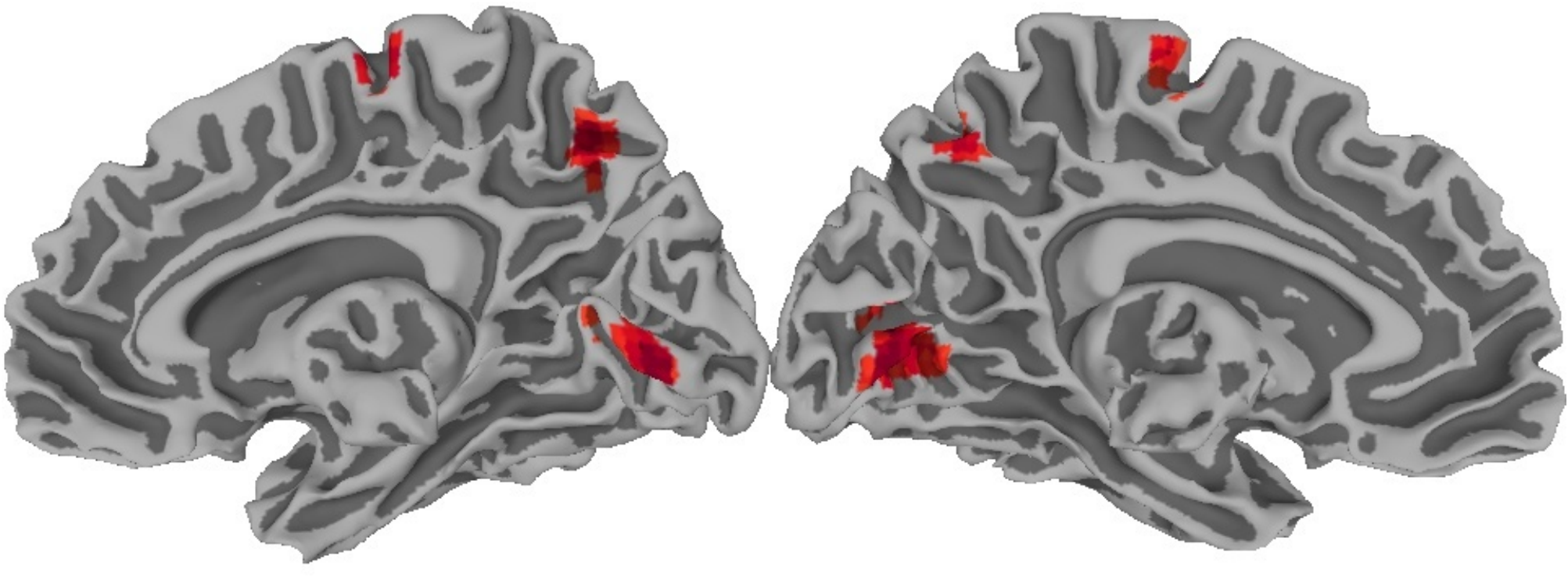}\label{fig:imagination-CTNN2}}

   \subfloat[]{\includegraphics[width=\linewidth]{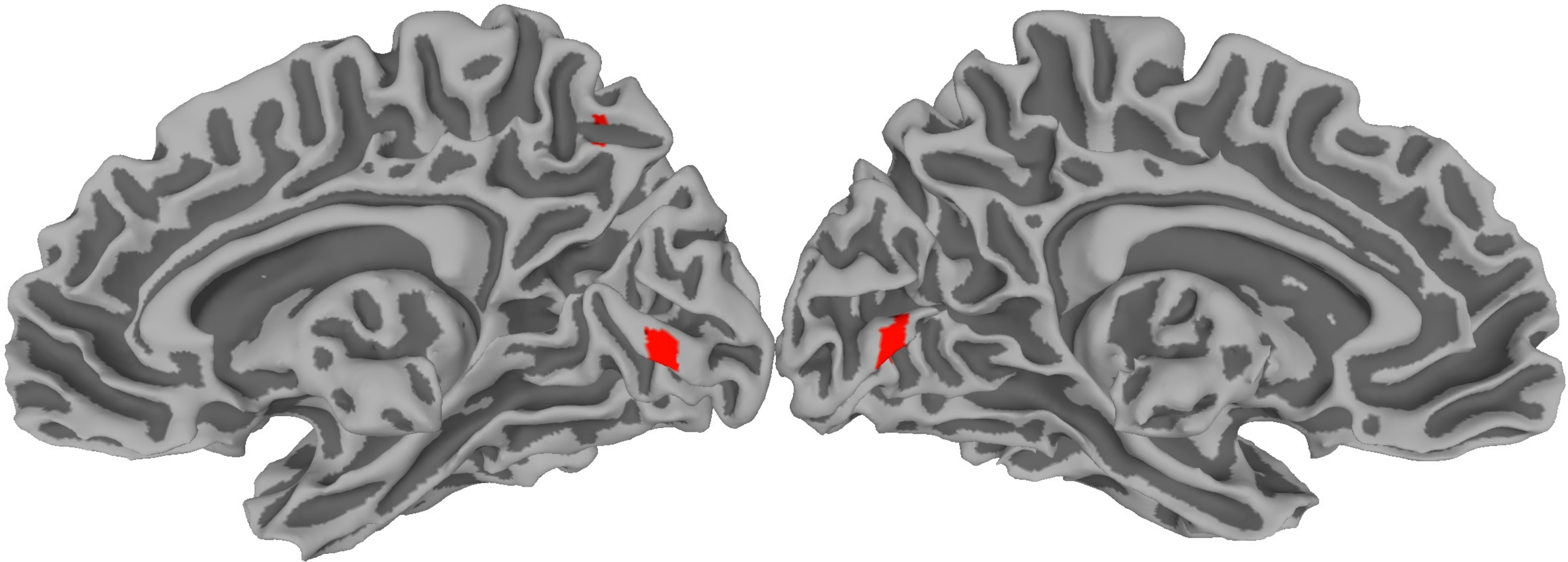}\label{fig:imagination-pTFCE}}

   \subfloat[]{\includegraphics[width=\linewidth]{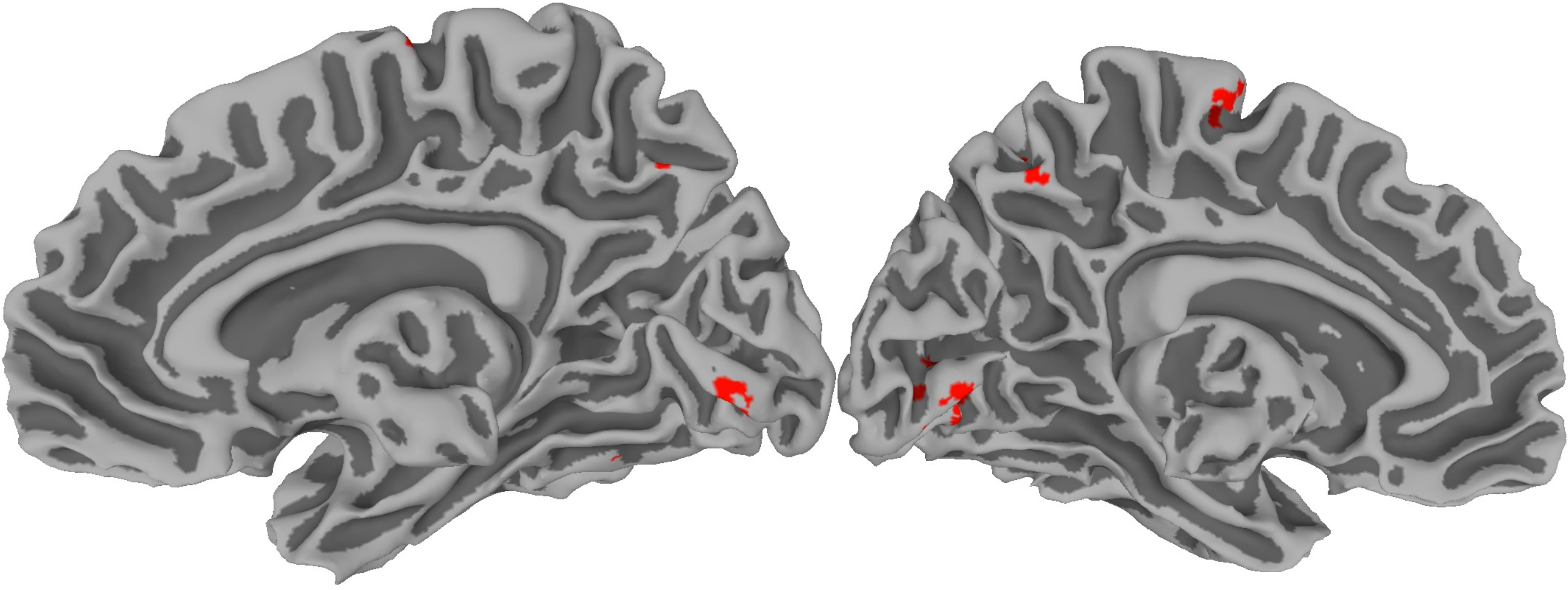}\label{fig:imagination-ARFAST}}

  \caption{Activation detected in healthy female volunteer performing
    a sports imagination experiment with (a) cluster-wise thresholding
    done using a second-order neighborhood, (b) pTFCE, and the (c) AR-FAST
    algorithm.}
  \label{fig:imag1}
 \end{figure}
 three types of neighborhood found  
 similar activated regions so we only display results
 (Fig.~\ref{fig:imagination-CTNN2})  using a second-order  neighborhood. The activation 
detected is mostly unemphatic. For instance, there is a very tiny blip of
 activation in the subject's 
 somatosensory cortex (
 left hemisphere) but not in her motor
cortex or the SMA. The precuneus and the primary visual cortex are identified,
but the extent of spatial activation (and clustering) is low. The
diffidence in the described activation map could have led to misinterpretations (if we were not,
for instance, aware of what we were expecting to be activated based on
our prior knowledge of the activity or on the fact that we had a
normal subject). It is important to note that in 
a clinical setting, where inference is desired at an individual
subject level it is often not known 
beforehand ({\em e.g.}, a patient with a TBI) the particular regions
of the brain that are expected to be activated.  


Unclearer activation maps are obtained by probabilistic Threshold-Free
Cluster Enhancement (pTFCE) \citep{spisaketal19} (Fig.~\ref{fig:imagination-pTFCE}) -- a refinement of Threshold-Free Cluster Enhancement (TFCE)
\citet{smithandnichols09} -- or by the robust 
fast adaptive and smoothed thresholding (AR-FAST) method of
\citet{almodovarandmaitra19} at $\alpha=0.05$
(Fig.~\ref{fig:imagination-ARFAST}). Activation is detected  barely in the precuneus, or (with AR-FAST) the primary motor cortex
(PMC). Other FAST methods \citep{almodovarandmaitra19} were also unsatisfactory.

This example indicates 
some of the challenges and
inadequacies associated with the most common  activation detection
methods. For this normal subject, activation in the SMA is
not easily detected by standard
thresholding methods, despite the fact that several studies have found 
this region to be reliably activated in healthy subjects during 
sports imagination experiments. These
deficits point to the need for improving
activation detection methodology, and we explore doing so by incorporating the additional
knowledge provided by both spatial context and prior belief on the
expected proportion of activated voxels. These two ingredients are
important and central to the application of activation detection in
fMRI and have hitherto never been successfully incorporated into the
statistical methodology for analysis. It is the goal of this paper to
redevelop such statistical analysis for the purpose of improving
activation detection in fMRI. 
We revisit this dataset in Section~\ref{sec:applications}
after developing our methodology.

\section{Methodological Development}
\label{mainsec:stat}
Sections~\ref{sec:background}-\ref{sec:merge} build the framework and
our methodology. The material can be highly technical, so we have
provided a  graphical overview in Section~\ref{sec:overview} for  
the reader who is most interested in a broad overview of our
contributions. 
\subsection{Background and Preliminaries}
\label{sec:background}
Most current approaches to the statistical analysis of fMRI data
involve post-hoc inference and analysis on obtained statistical parametric maps (SPMs). For instance, a popular approach
to incorporating spatial context in activation detection is to use cluster-wise
thresholding, or its derivatives, on the estimated $p$-vaues of test statistics obtained
after fitting an appropriate model at each voxel. Another important
aspect of fMRI activation, that is, the {\em a
  priori} known low expected proportion of active voxels is largely ignored. In this section, we
address ways to formally incorporate these two features of fMRI
activation in the statistical modeling and analysis. 

As in most current software, we let $\bp = (p_1,p_2,\ldots,p_n)$ be
the $p$-values of the test statistic at the $n$ voxels in a fMRI
volume, after fitting an appropriate model to test for activation in a first-stage analysis. We emphasize that this
approach is marginally applied at each voxel for computational
considerations, and also that the resulting statistics and their
estimated $p$-values still maintain their dependencies that we account
for shortly. Now, if the voxel-wise test
statistic is known, for example, to be $t$-distributed, then each $p_i$ can be
modeled  by an exact mixture model~\citep{maitra09b}. However, we
instead build our method using an  approximate but more general
framework~\citep{parkerandrothenberg88,allisonetal02,poundsandmorris03} 
that models  each $p_i$ as a mixture of the uniform (for the
inactivated voxels) and beta distributions (for the different kinds,
in terms of intensity of activation, of activated voxels).
Let $b(\cdot;\alpha_k,\beta_k)$ be the beta density
with shape parameters $(\alpha_k,\beta_k)$. Then
\begin{equation}
p_i\sim f(p_i;\bpi,\balpha,\bbeta) = \pi_0 + \sum_{k=1}^K\pi_kb(p_i;\alpha_k,\beta_k).
\label{obsllhd}
\end{equation}
with 
constraints:
\begin{enumerate}
\item[(i)]
  \label{constr1}
 $\bpi=(\pi_0,\pi_1,\ldots,\pi_K)^\top$ lies on the  $K$-dimensional simplex,
\item[(ii)]  \label{constr2}
  $\pi_0\geq\delta$ and
  \item[(iii)]   \label{constr3}
$\alpha_k < 1 < \beta_k$ and
$\alpha_k/(\alpha_k+\beta_k) \leq \eta$  $\forall$ $1\leq k\leq 
K$ with $\eta$  the mean of the distribution of the $p$-value from
an activated component. 
\end{enumerate}
From (ii), $\delta$ is  the  {\em a   priori} expected minimum
proportion of inactivated voxels while 
(iii) 
expresses that $p_i$s from activated components are expected to be declared
significant.     

The generative model for
\eqref{obsllhd} stems from the distribution of 
the (observed) $p$-value at each voxel along with its (missing) class indicator. Let
$W_{ik}$ be 1 if the $i$th voxel is in the
$k$th class ($k=0,1,\ldots,K$) and 0 otherwise. Writing
 $\balpha = (\alpha_1,\alpha_2,\ldots,\alpha_K)^\top$, 
$\bbeta=(\beta_1,\beta_2,\ldots,\beta_K)^\top$ and $\bW =
((W_{ik}))_{1\leq i\leq n;0\leq k\leq K}$,  the complete log
likelihood, assuming that the dependence in the $\bp$s is through the 
generative $\bW$s, is 
\begin{equation}
  \begin{split}
    \ell_c(\bpi,\balpha,\bbeta;\bW,\bp) = & \sum_{i=1}^n 
    W_{i0}\log\pi_0 \\ & +
     \sum_{i=1}^n\sum_{k=1}^KW_{ik}[\log\pi_k  +\log b(p_i;\alpha_k,\beta_k)].
   \end{split}
  \label{compllhd}
\end{equation}
The Expectation-Maximization (EM) algorithm~\citep{dempsteretal77,MclachlanandKrishnan08} 
iteratively maximizes
$\E[\ell_c(\bpi,\balpha,\bbeta;\allowbreak\bW,\bp)\mid \bp]$ 
to yield maximum likelihood estimates (MLEs) of
$(\bpi,\balpha,\bbeta)$.

\subsection{Incorporating spatial context}
\label{sec:practical}
A shortcoming of \eqref{compllhd},  and thus of \eqref{obsllhd},
is its disregard of  spatial context
. 
Modeling the $W_{ik}$s via a Markov Random Field ({\em e.g.} Potts' model) 
is computationally impractical because the expectation step (E-step)
expressions are intractable, computed at each iteration only through
time-consuming  Markov Chain Monte Carlo (MCMC) or other methods
. Our solution is to penalize
\eqref{compllhd} by adding the term
 \begin{equation}
 - \sum_{i\neq j =1}^n\sum_{k=0}^K
W_{ik}W_{jk}\sum_{d=1}^3\left[\frac{(v_{id}  -
    v_{jd})^2}{h_{kd}^2} -  \log h_{kd}\right],
\label{penalty}
\end{equation}
with $v_{id}$s as the 
the $i$th voxel's coordinates
in the 3D image volume 
($v_{i3}$ is omitted for 2D slices). For given $k$, \eqref{penalty}
has the least effect  when $W_{ik} W_{jk} = 1$  for nearby
voxels  $i$ and $j$, and is like 
a regularization term in density  estimation (but note that the $v_{id}$s are
  fixed here, while $W_{ik}$s are unobserved). The $h_{kd}^2$s in the
  denominator   modulate the influence of 
  \eqref{penalty}, varying by axes and with $k$ to allow for
  localization. The additional $\log h_{kd}$ terms are for
  computational convenience (ensuring also that \eqref{penalty}
  is, but for a constant term of addition, a log density). 

  The terms for each axis ($d$) in \eqref{penalty} can be recast as
  \begin{equation*}
\sum_{i\neq j =1}^n\sum_{k=0}^K W_{ik}W_{jk}\frac{(v_{id} -
  v_{jd})^2}{h_{kd}^2} = \sum_{i=1}^I\sum_{k=0}^K
W_{ik} \frac{(v_{id}-\mu_{kd})^2}{\sigma_{kd}^2}.
\end{equation*}
where we recast $\mu_{kd}\doteq{\sum_{i=1}^nW_{ik}v_{id}}/{\sum_{i=1}^nW_{ik}}$ and
$\sigma_{kd}^2\doteq h^2_{kd}/2$ for $d=1,2,3$. The penalized complete
log likelihood (\eqref{compllhd} + \eqref{penalty}) is
\begin{equation*}
  \begin{split}
\sum_{i=1}^n\sum_{k=0}^K  W_{ik} \left[\log\right.& \pi_k  + \log b(p_i;\alpha_k,\beta_k) \\ & \left. -\log|\bSigma_k|/2-(\bv_i-\bmu_k)^\top\bSigma_k^{-1}(\bv_i-\bmu_k)/2\right].\end{split}
\end{equation*}
where $\alpha_0=\beta_0\equiv 1$, $\bmu_k =
(\mu_{k1},\mu_{k2},\mu_{k3})^\top$,
$\bv_i=(v_{i1},v_{i2},v_{i3})^\top$ and $\bSigma_k =
diag(\sigma_{k1}^2,\sigma_{k2}^2,\sigma_{k3}^2)$.
Maximizing~\eqref{obsllhd} after incorporating the spatial penalty
term is equivalent to maximizing
$  \ell(\bTheta;\bp,\bv) =\sum_{i=1}^n\log\left\{\sum_{k=0}^K\pi_kf(p_i, \bv_i; \bGamma_k)\right\}$
with $f(p_i, \bv_i; \bGamma_k) =
b(p_i;\alpha_k,\beta_k)\phi(\bv_i;\bmu_k,\bSigma_k)$, where 
$\bTheta=\{\bTheta_0,\bTheta_1,\ldots,\bTheta_K\}$ with $\bTheta_k =
\{\pi_k,\bGamma_k\}$, $\bGamma_k=\{\alpha_k,\beta_k,\bmu_k,\bSigma_k\}$,
$\phi(\bx;\bmu,\bSigma)$ is  
the multivariate normal density with parameters $\bmu$, $\bSigma$,
and $(\bpi, \balpha,\bbeta)$ has the same  constraints as
\eqref{obsllhd} ($\alpha_0=\beta_0\equiv 1$ and do
not have to be estimated). From {\em an optimization perspective},
this maximization 
is essentially equivalent to finding MLEs in a mixture model where each
component $f(p_i,\bv_i;\bGamma_k)$ is the product of a
beta density $b(p_i;\alpha_k,\beta_k)$ and a
multivariate 
Gaussian density  $\phi(\bv_i;\bmu_k,\bSigma_k)$
with diagonal $\bSigma_k$ 
and mixing proportion $\pi_k$, for $k=0,1,\ldots,K$. For a given $K$,
the number of parameters to be estimated is $d_K=K(3 + 2c_v)+2c_v$ with $c_v=2$ or
$3$ for a 2D slice or 3D volume, respectively.

\subsection{Parameter estimation and model-fitting}
\label{sec:estimation}
Adding~\eqref{compllhd} and~\eqref{penalty} yields the penalized complete-data
loglikelihood 
\begin{equation}
  \begin{split}
    \sum_{i=1}^n\sum_{k=0}^K  W_{ik}\bigg\{\log& \pi_k +
    \log b(p_i;\alpha_k,\beta_k)
    \\&
    -\frac12\log|\bSigma_k|
      - \frac12(\bv_i-\bmu_k)^\top\bSigma_k^{-1}(\bv_i-\bmu_k) \bigg\} 
  \end{split}
    \label{modcompllhd}
\end{equation}
while the penalized observed-data log likelihood equation is
\begin{equation}
  \begin{split}
  \ell(\bTheta;\bp,\bv) &=\sum_{i=1}^n\log\left\{\sum_{k=0}^K\pi_k
    f(p_i, \bv_i; \bGamma_k)\right\}
\\&
  \equiv  \sum_{i=1}^n\log\left\{\sum_{k=0}^K\pi_kb(p_i;\alpha_k,\beta_k)\phi(\bv_i;\bmu_k,\bSigma_k)\right\}.
\end{split}
\label{modobsllhd}
\end{equation}

Development of a classical EM algorithm for parameter estimation given
$K$ is  almost immediate, but 
the inseparability of 
$(\alpha_k,\beta_k)$s from the beta function, the large scale 
of the problem with hundreds of thousands of voxels (in 3D), the
need for constrained optimization and repeated 
initialization call for many 
refinements~\citep{lange10,wu83,boydandvandenberghe04,mengandvandyk97}. 
We develop strategies for parameter estimation and provide custom
initialization schemes and a novel parallel  
Expectation-Gathering-Maximization (EGM) method for  implementing
the Alternating
Partial Expectation Conditional Maximization (APECM)
algorithm~\citep{chenandmaitra11}
through its  APECMa variant~\citep{Chen2013}.

\subsubsection{Parameter estimation using the EM algorithm}
\label{sec:EM}
The classical EM algorithm has the E-step for calculating $
w_{ik}^{(s + 1)}
  = \E[W_{ik} \mid \bp,\bv, \bTheta^{(s)}]
  =
  {
    \pi_k^{(s)} f(p_i,\bv_i ; \bGamma_k^{(s)})
  }/{
    \sum_{j = 0}^K \pi_j^{(s)} f(p_i,\bv_i ; \bGamma_j^{(s)})
  }
  $ at the $(s + 1)$th iteration, with $\bTheta^{(s)}$ being the current
  value of $\bTheta$  obtained  
  at the end of the $s$th iteration. At that iteration, the
  maximization step (M-step)
  maximizes 
  \begin{equation}
    \label{eqn:Q}
    \begin{split}
  Q(\bTheta; \bTheta^{(s)})&
  = \E[\ell_c(\bTheta, \bW \mid \bp,\bv) ; \bTheta^{(s)}]
\\&
  =
  \sum_{i = 1}^n \sum_{k = 0}^K w_{ik}^{(s + 1)} 
  [ \log \pi_k + \log f(p_i,\bv_i; \bGamma_k) ]
  \end{split}
  \end{equation}
  within the
  constraints~\eqref{constr1},~\eqref{constr2}
  and~\eqref{constr3} that govern~\eqref{obsllhd}.
  The maximization of~\eqref{eqn:Q} is   separable in  $\bmu_k$s,
  $\bSigma_k$s, $\bpi$ and  $(\balpha,\bbeta)$ and can be handled one
  subset of parameters at a   time.  M-step updates for $\bmu_k$ and 
  $\bSigma_k$ are immediate, with $\hat\bmu_k^{(s+1)} =
  \sum_{i=1}^nw_{ik}^{(s+1)}\bv_{i}/\sum_{i=1}^nw_{ik}^{(s+1)}$ and the
  (diagonal) matrix $\hat\bSigma_k^{(s+1)}$ containing the diagonal
  elements of
  $\sum_{i=1}^nw_{ik}^{(s+1)}(\bv_{i}-\hat\bmu_{k}^{(s+1)})(\bv_{i}-\hat\bmu_{k}^{(s+1)})^\top/\sum_{i=1}^nw_{ik}^{(s+1)}$. Updating
  $\bpi$ under
  constraints~\eqref{constr1} and~\eqref{constr2} requires some care however, so
  we state and prove the following
  \begin{result}
    \label{theo:result}
    Let $g(\bpi;\bW) = \sum_{i=1}^n\sum_{k=0}^Kw_{ik}\log\pi_k$,  with
    $\bpi=(\pi_0,\pi_1,\ldots,\allowbreak\pi_K)^\top$. Suppose that
    \begin{enumerate}
  \item[(a)]\label{constra} $\bpi^\ast$
    maximizes $g(\bpi;\bW)$ under the constraint that $h_e(\bpi)=\bpi^\top\bone-1=0$
    where $\bone = (1,1,\ldots,1)^\top$ inside the convex set
    $\bPi=\{\bpi:0\leq\pi_k\leq1,k=0,1,\ldots,K\}$.
\end{enumerate}
    Then,   under the additional 
    inequality constraint
    \begin{enumerate}
    \item[(b)]\label{constrb} $h_\iota(\bpi) = \pi_0 - \delta \geq 0$,
    \end{enumerate}
      $g(\bpi;\bW)$ is maximized at $\pi^\bullet_0 =
    \max{(\delta,\pi_0^\ast)}$ and $\pi^\bullet_k =
    (1-\pi_0^\bullet)\pi_k^\ast/(1-\pi_0^\ast)$ for all
    $k=1,2,\ldots,K$.   
  \end{result}
  \begin{proof}
    See~Appendix~\ref{proof}.
  \end{proof}

Result~\ref{theo:result} means that we can first ignore the inequality
constraint~(b) and estimate $\bpi$  only under the equality constraint
(a) to get $\hat\pi_{k\ast}^{(s+1)} = 
    \sum_{i=1}^nw_{ik}^{(s+1)} /
        {\sum_{k=0}^K\sum_{i=1}^nw_{ik}^{(s+1)}}$.
   Then, if $\pi_{0\ast}^{(s+1)} \geq \delta$, then
  $\hat\pi_k^{(s+1)}\equiv \hat\pi_{k\ast}^{(s+1)}$ for all
  $k=0,1,\ldots,K$. Otherwise, set $\hat\pi_0^{(s+1)} \equiv \delta$
  and $\hat\pi_k^{(s+1)} =\frac{
  (1-\delta)\sum_{i=1}^nw_{ik}^{(s+1)}}{\sum_{k=1}^K\sum_{i=1}^nw_{ik}^{(s+1)}}$
  for all $k=1,2,\ldots,K$.

  Since we do not have analytical expressions for M-step updates of
  $\balpha$ and   $\bbeta$ within the constraints imposed by (iii), we
  use constrained optimization subject to linear inequality
  constraints and using an adaptive barrier, as in the
  \code{constrOptim()} function~\citep{lange10} 
  in \proglang{R}~\citep{Rcore}. Here also, $(\alpha_k,\beta_k)$ can be
  addressed separately for each $k=1,2,\ldots,K$. The
  constrained optimization for $\alpha_k$ and $\beta_k$ may not
  result in convergence at each M-step iteration owing to numerical
  issues. However, an increase in~\eqref{eqn:Q} is enough to guarantee 
  convergence of the ML estimates for \eqref{modobsllhd} as per the
  generalized EM algorithm~\citep{wu83}. Therefore, for
  $\balpha$ and $\bbeta$, we ensure an increased \eqref{eqn:Q} 
  within each M-step.
  We use the relative increase in the value of ~\eqref{modobsllhd}
  over successive iterations
$
[ \ell(\bTheta^{(s + 1)} ; \bp, \bv) - \ell(\bTheta^{(s)} ; \bp,\bv)]
 / |\ell(\bTheta^{(s)} ; \bp,\bv)|
 < \epsilon
$
to determine convergence of our EM algorithms. In this paper, we set $\epsilon = 10^{-6}$.

\subsubsection{Initialization}
\label{sec:initialization}
The EM algorithm only guarantees convergence to a local 
maximum with the initializing parameter values greatly influencing the
converged ML estimates~\citep{maitra09a}. The need for effective
initialization becomes more acute with increasing dimensionality and
$K$~\citep{maitraandmelnykov10}. Therefore, the choice of initial values
is an important determinant in EM's performance. In order to
increase the possibility of convergence to a global maximum, 
we modify the {\it Rnd-EM}
approach of~\citet{maitra09a} to initialize $\bTheta^{(0)}$ in the
presence of constraints. Exact specifics on the choice of the random
initializing seeds are in Appendix~\ref{sec:init}. 
\subsubsection{Parallelization of computations}
\label{sec:parallel}

Despite its general wide appeal, the
EM algorithm of Section~\ref{sec:EM} is plagued by
slow convergence and can be time-consuming for typical 3D
fMRI datasets with potentially upto a million voxels. While many speedup
methods~\citep{MclachlanandKrishnan08} exist, we
incorporate the  Alternating Partial Expectation Conditional
Mazimization  (APECM, APECMa) algorithms of \citet{chenandmaitra11}
and \citet{Chen2013}
that utilize extra space to store and reduce calculations while being 
easily parallelizable on multi-core multi-processor computing
architectures. Specifically, APECM and APECMa modify the
Alternating Expectation Conditional
Maximization~(AECM) algorithm of \cite{mengandvandyk97} for mixture models
by introducing a partial calculation of the E-step from the previous 
iteration (denoted as PE-step). This 
parsimonious computing yields faster EM convergence rates and shorter computing times. Further, data-distributed algorithms, such as the Expectation Gathering Maximization~(EGM) algorithm~\citep{Chen2013}
for distributed computing architectures  can be coupled with APECM and
APECMa to  parallelize the M-step (really the conditional M- or
CM-step) computations when local sufficient statistics are
available. We refer to Appendix~\ref{sec:APECM} for the exact details of our
implementation of EGM with APECMa in our context.
\subsubsection{Choice of $K$}

Many methods~(see {\em e.g.} \citep{Melnykov2010}) exist for assessing $K$ in a mixture
model. The most popular are Akaike's An (AIC)~\citep{akaike73} and  
Bayes (BIC)~\citep{schwarz78} Information Criteria (respectively,
selecting $K$ to be the one minimizing $\mB_K=-2 
\ell(\hat{\bTheta_K};\bp,\bv) + 2d_K$. 
Here  $\hat\bTheta_K = \{(\hat\pi_k,\hat\bGamma_k): k = 0, 1, \ldots, K\}$
contains MLEs of $\bTheta$ obtained for a given
$K$, and $d_K$ is the number of unrestricted parameters to be
estimated in the model. Our approach to choosing $K$ uses BIC, but additionally
incorporates \citet{kassandraftery95}'s suggestion that only
reductions of BIC beyond 10 providing very strong evidence in favor of
the more parameter-rich (larger-$K$) model. Formally, we choose the
number of components as the first $K$ for which $\mB_K \leq
\mB_{K+1}+10$, for $K=0,1,\ldots,K_{\max}-1$.  We set $K=K_{\max}$ if
no $K$ satisfies this condition.
\subsection{Merging of components and voxel classification}
\label{sec:merge}
\subsubsection{Merging with inactive component}
Having determined $K$, 
we 
obtain an initial {\em maximum   a posteriori} (MAP) 
classification of each voxel using the mixture model and estimated MLEs. 
Constraint~\eqref{constr3} of~\eqref{obsllhd} on the activated beta components  ideally   
means that for $k > 0$, we get beta components with
parameter estimates
$(\hat\alpha_k,\hat\beta_k)$ 
$\hat\alpha_k/(\hat\alpha_k+\hat\beta_k)
\leq \eta$. This last constraint on the
expectation of the (activated) beta
components logically follows from
our expectation that these
$p$-values are low for the activated
components. We use $\eta=0.05$ to
match the most common threshold used
in declaring an alternative hypothesis
significant. 
However,  the
spatial penalty can also yield beta  components 
with true mean $\alpha_k/(\alpha_k+\beta_k) \geq\eta$, 
because the spatial context does not involve 
the constraint~\eqref{constr3} or the observed $p$-value. We merge
such components with the 
inactivated (zeroeth) component via a hypothesis
test of constraint~\eqref{constr3}.   Specifically, for the $k$th 
($k=1,2,\ldots,K$) estimated beta component, we test  
$H_0: {\alpha_k}/({\alpha_k + \beta_k}) \geq \eta$ against
$H_a: {\alpha_k}/({\alpha_k + \beta_k}) < \eta$. We use
the likelihood ratio test statistic (LRTS)
$\Lambda_k = -2 \sum_{i=1}^{n_k} \{\log{[    b(p_{ki};
    \hat{\alpha}_{0,k}, \hat{\beta}_{0,k})]} - \log{[b(p_{ki};
    \hat{\alpha}_k, \hat{\beta}_k)]}\}$ where $n_k$ is the
number of voxels classified to the $k$th component, 
$p_{ki}$s are the $p$-values of those voxels, $\hat{\alpha}_{0k}$ and
$\hat{\beta}_{0k}$ are MLEs under $H_0$ and
$\hat{\alpha}_k$ and $\hat{\beta}_k$ are MLEs under $H_0\cup H_a$.
The $p$-value of each of these $K$ LRTSs is
calculated using $\Lambda_k \sim \chi^2_1$, then converted to
$q$-values to  control for false
discoveries~\citep{benjaminiandhochberg95}. 
Components for which we fail to reject $H_0$ 
are 
merged with the inactivated
component.
We write $\hat K$ as  the number of activated comments remaining after these mergers.

\subsubsection{Merging of active components}
Similar to the case of 
inactivated components being possibly split into multiple groups because of
spatial constraints, some of the
remaining activated components are
possibly identified as distinct
only because they are spatially
disjoint. We distinguish between 
spatially disjoint pairs of activated components with similar effect
sizes from those that have distinct effect sizes. For
each pair of activated components, we compared the parameters of the 
beta distributions representing the $p$-values of the
components. Specifically, we tested the $k$th and $l$th activated component
means using  $H_0: (\alpha_k,\beta_k) = (\alpha_l, \beta_l)$
against 
the alternative $H_a: (\alpha_k,\beta_k) \neq (\alpha_l, \beta_l)$
using a LRTS which is given by
$\Lambda_{kl} = \sum_{i=1}^{n_k} \{\log{\{b(p_{ki};
  \hat{\alpha}_{kl}, \hat{\beta}_{kl})\}}+\sum_{i=1}^{n_l} \log{\{b(p_{li};
  \hat{\alpha}_{kl}, \hat{\beta}_{kl})\}} -\sum_{i=1}^{n_l}
\log{\{b(p_{ki}; \hat{\alpha}_k, \hat{\beta}_k)\}}
-
\sum_{i=1}^{n_l}\log{\{b(p_{li}; \hat{\alpha}_l, \hat{\beta}_l)\}}$
where $p_{ri},i=1,2,\ldots,n_r$ is the set of $p$-values at
the $n_r$ voxels classified to the $r$th component, $r\in\{k,l\}$, 
$(\hat\alpha_r, \hat\beta_r)$ is the MLE of $(\alpha_r,\beta_r)$
obtained from fitting a beta distribution to the $p$-values classified
to the $r$th component, and $\alpha_{kl},\beta_{kl}$ are the parameter
MLEs obtained upon fitting a beta distribution to the combined sample
of $p$-values.  Under the null hypothesis, the asymptotic distribution
of $-2\mbox{LRTS}\sim\chi_2^2$. To account for 
the $\hat{K}\choose2$ pairwise tests, we used $q$-values~\citep{benjaminiandhochberg95} 
larger than 0.05 to merge pairs of spatially separated active clusters
with indistinguishable beta distribution parameters.


\subsubsection{Voxel classification}
\label{sec:voxel}
Voxels are initially classified into the initial $\hat K$ groups in
terms of the highest MAP. After the potential mergers of
Section~\ref{sec:merge}, the voxels in the merged groups provide a final classification of each voxel as inactivated or
activated (in one of the merged, activated groups). The (merged)
activated groups represent activated voxels of different intensities
and correspond to different strengths of activation.

\subsection{Overview of our methodology}
\label{sec:overview}
\begin{figure*}[h]
  \includegraphics[width=\textwidth]{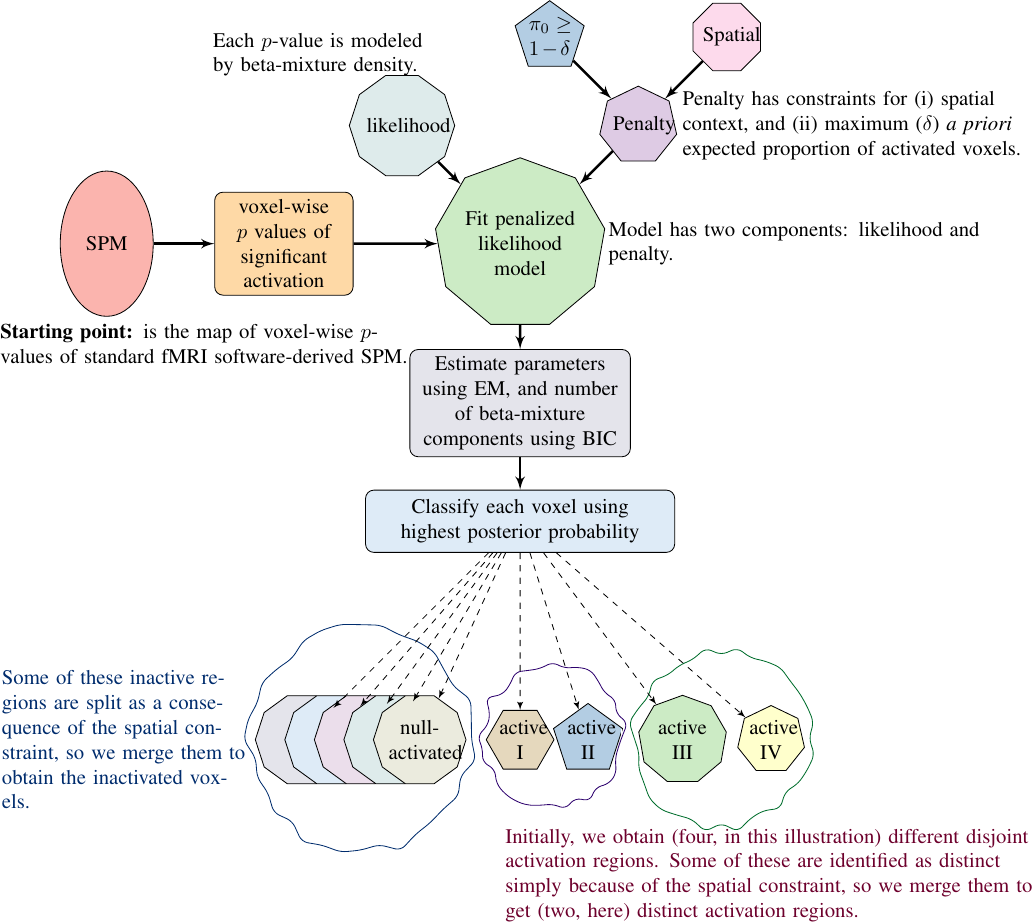}
\caption{Graphical illustration of our modeling,    estimation and
 algorithmic framework. Abbreviations used: SPM, Statistical
 Parametric Map; EM, Expectation
 Maximization algorithm; BIC: Bayesian Information Criterion.} 
\label{fig:overview}
\end{figure*}
Figure~\ref{fig:overview} provides a flowchart of all the  steps  in
our methodology. The starting point is the map of voxel-wise
$p$-values of a SPM obtained from some standard fMRI processing
software. Each $p$-value is modeled by a mixture density, as
in~\eqref{obsllhd}. To the loglikelihood model derived from these
mixture densities, we add penalties for spatial context and
constraints for the maximum {\em a priori} expected proportion of
activated voxels, as detailed in Sections~\ref{sec:background}
and~\ref{sec:practical}. We then use the development in
Section~\ref{sec:estimation} to estimate the parameters of the beta
mixture model, and the number of mixture components (using BIC). Each
voxel is classified to the mixture component having the highest 
posterior probability. The components are then merged, per the results
of hypothesis tests, following the development of Setion~\ref{sec:merge}.

\subsection{Extension to a two-sided testing framework}
\label{sec:2sided}
Our methodology is easily extended to  situations involving
two-sided $p$-values. In this situation, we consider the two one-sided
$p$-values (for the left and right tail) individually, and run our
methodology, with $\delta$ replaced in each case by $\delta_l$ and
$\delta_r$ depending on the practitioner's minimum {\em a priori} expected
proportion of inactive voxels in the right and left tail. If the
practitioner can only provide an overall minimum {\em a priori}
expected proportion of inactive voxels ($\delta$), then we suggest
using $\delta_l=\delta_r=1 - (1-\delta)/2$. We illustrate and evaluate
performance using the above methodology on 26 Flanker task datasets
in Section~\ref{sec:flanker}.

In this section, we have developed and fine-tuned statistical
methodology for application to detecting activation in fMRI
studies. We now evaluate its performance in realistic simulation
experiments before applying it to our sports imagination experiment
dataset. 
\section{Performance Evaluations}
\label{sec:simulations}
We very comprehensively evaluated our activation detection
methodology. Our evaluation setups included
simulation experiments in realistic 2D and 3D settings, and on 81 real
datasets from four different task paradigms. The performance of our
activation detection methodology (denoted as MixfMRI or Mf in our
figures) was compared with a host of commonly used
 alternatives, that either have software available or
are easily coded, and, because our evaluations are over hundreds of
simulated and real datasets, are computationally practical to use,
implement and evaluate. Consequently, we do not include comparisons with
Bayesian methods, such as those in~\citet{smithandfahrmeir07} that use
computationally expensive MCMC methods and were
demonstrated only in 2D, and for which code is unavailable.
Our comparisons only  use 
the enhanced versions of methods if the improvements have been shown
to improve activation detection: for instance, we use pTFCE instead of
TFCE~\citep{spisaketal19}, and the fully-automated FAST
methods~\citep{almodovarandmaitra19} instead of structural
adaptive segmentation~\citep{polzehletal10}. An added benefit here is
the ready availability of software for both pTFCE and FAST (while the
R package \pkg{fmri} implements structural adaptive segmentation,
the software when applied to our datasets threw up  errors that  were
not resolved even after communicating with the 
authors). We note also that \citet{almodovarandmaitra19} showed
poorer performance of both TFCE~\citep{smithandnichols09} and adaptive
segmentation~\citep{polzehletal10} in their extensive simulation and real
data experiments.  Therefore, for comparison, we use (a) thresholding,
after controlling for false discoveries
\citep{benjaminiandyekutieli01} at 0.05 (abbreviated as FDR in 
our figures), (b) thresholding computed by using Random Field theory
(abbreviated using RF)
 to get a specified $p$-value of 0.05~\citep{worsley94}, (c)
cluster-based thresholding with a first-order (CT-1st), second-order (CT-2nd)
or (and in the case of 3D) third-order (CT-3rd) neighborhoods, all
with cluster sizes decided by AFNI and a thresholded $p$-value of
0.001, following \citet{wooetal14}, (d) the three sets of FAST
methods~\citep{almodovarandmaitra19} with parameter $\alpha =0.01$ or
$0.05$: AM-FAST (AM, $\alpha=0.01$ or AM, $\alpha=0.05$ in
our figures, AR-FAST (AR, $\alpha=0.01$ or AR, $\alpha=0.05$), and
ALL-FAST (ALL, $\alpha=0.01$ or ALL, $\alpha=0.05$), (e) permutation
testing at $\alpha=0.05$ (``Permutation'' in our
figures), and (f) the spatial mixture model method (or SMM in our figures), 
of~\citet{hartvigandjensen00}. We also evaluated performance, in the
case of our 2D experiments, using our methodology that included the
proportion-of-inactives ($\pi_0\geq \delta$) constraint, but not the spatial 
constraint (denoted as MfnoX in our figures). Therefore, including the
methodology developed here, and 
with different specifications of $\delta$, our 
comparisons are over a total of 19 methods. 
Our methodology, with or without spatial constraints, was applied
using our \proglang{R} package~\pkg{MixfMRI}~\citep{Chen2017}, while
we used our coded \proglang{R} 
scripts for FDR and permutation testing. The \proglang{R} package
\pkg{AnalyzeFMRI}~\citep{Bordier2011} was used for RF, cluster-wise
thresholding and spatial mixture modeling, while we used
\pkg{pTFCE}~\citep{spisaketal19} for pTFCE, and
\pkg{RFASTfMRI}~\citep{almodovarandmaitra19} for the FAST
methods.
For all methods, activation detection accuracy was numerically assessed 
by the Jaccard index~$\mJ$~\citep{jaccard1901,maitra10} that is the ratio of
the number of voxels that are both identified as active and truly so
and the number of voxels that are either truly active, so detected or 
both. The index $\mJ$  takes values in [0,1], with 
0 indicating that no voxels are correctly identified as activated and
1 indicating perfect detection. Finally, we evaluated the
unique ability of MixfMRI to detect the different kinds of
activation by calculating the adjusted Rand
index~\citep{hubertandarabie85} between our estimated activation
maps and the ground truth in our simulation experiments. This index,
or ARI, takes values in $(-\infty,1]$ with  the upper limit
indicating perfect predicted classification, and values farther away
from 1 signifying poorer agreement of the classification with the
truth. The ARI is expected to be zero when the voxels
have been randomly classified.
\subsection{Simulation Experiments}
\label{sec:simulation}
\subsubsection{Two-dimensional framework}
\label{sec:2d}
\begin{figure}[h]
  \mbox{
    \hspace{-0.01\textwidth}
    \subfloat[$\pi_0=0.991$]{\includegraphics[width=.33\linewidth]{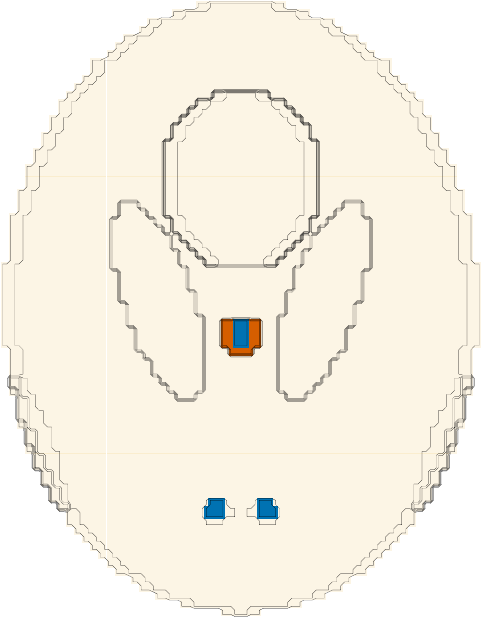}}
    \hspace{-0.0\textwidth}
    \subfloat[$\pi_0=977$]{\includegraphics[width=.33\linewidth]{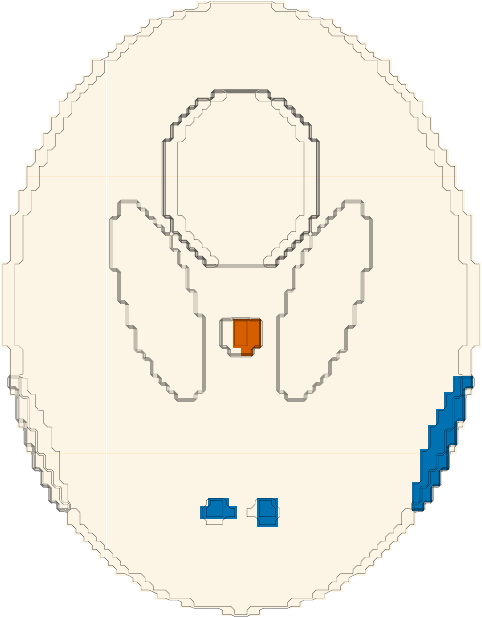}}
    \hspace{-0.0\textwidth}
    \subfloat[$\pi_0=0.96$]{\includegraphics[width=.33\linewidth]{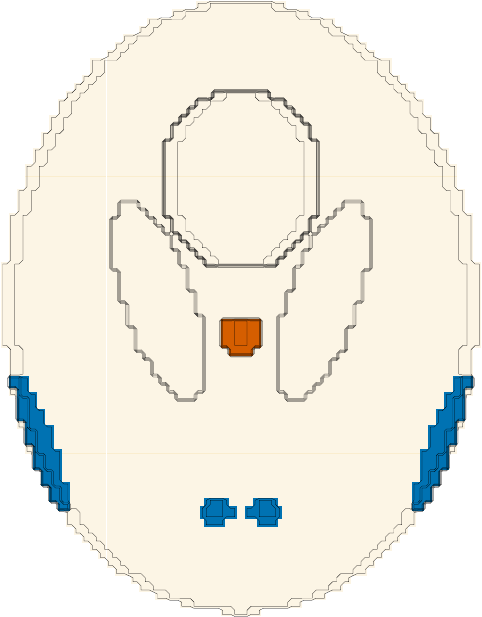}}}
  \caption{
    The phantom with three areas of simulated activation (in darker colors) that was used
    in our experiments. (The two dark colors indicate the two
    different intensities of activation.) In (a), the phantom has very few
    areas of true activation while in case (b), the phantom shows
    activation only   on one side (and with smaller spatial extent)
    while in (c) the phantom shows activation that is bilateral (on both sides).
  } 
  \label{fig:phantom}
\end{figure}

\paragraph{Experimental setup}
\label{sec:experimental}
We evaluated our methodology on digitized 2D
phantom~\citep{vardietal85,maitraandosullivan98} datasets commonly
used in Positron Emission Tomography experiments and specifically
redesigned by us to mimic  $p$-values typically obtained in fMRI
experiments.
 Our setup is in 2D to permit a very  thorough and detailed large-scale
analysis, while also being realistic enough to capture the challenges
in  3D settings. Our phantom (Figure~\ref{fig:phantom}) had
different-sized and oriented digitized ellipses, each representing
hypothesized structures in the brain with some of them presumed activated by
stimulation. We considered three scenarios.
 The first case (Figure~\ref{fig:phantom}a) had very little  (0.9\%
of all ``in-brain'' pixels) true activation. The second case 
(Figure~\ref{fig:phantom}b)  had 2.3\% truly
activated in-brain pixels,  with all activation  on one side of the
brain, as would happen, say, in a one-hand finger-tapping
experiment. The third case  
(Figure~\ref{fig:phantom}c) had 3.97\% of   truly
activated in-brain pixels with activation that is symmetric on both sides (as 
expected, say, in a bilateral finger-tapping  
experiment.) In each case, two regions (shown in darker colors) of differing
strength are truly activated.
Thus, the true $\bpi$ is (0.991, 0.005, 0.004), (0.977, 0.019, 0.004) or
(0.96, 0.034, 0.006) for the scenarios in Figure~\ref{fig:phantom}. For
all cases,  $K=2$ is the number of active  components when
ignoring spatial context. However the activated components span multiple
homogeneous spatial regions. 
Ignoring spatial context, each of the two activated components
corresponds to one of the  two
non-uniformly distributed components of our generative mixture model
whose choice we now discuss.

For simulating the $p$-values for our experiments, we forwent the
beta distributions for the $p$-values of the  activated
regions, and instead
chose the distribution of the $p$-value
of a one-sided normal test statistic: $\psi(p;\nu_k) =
\phi(\Phi_z^{-1}(1-p);\nu_k,1)/\phi_z(\Phi_z^{-1}(1-p))$, where
$\Phi_z(\cdot)$ and $\phi_z(\cdot)$ are the standard normal
distribution and density functions, and $\phi(\cdot;\nu_k,1)$ is the
normal density function with mean $\nu_k$ and variance $1$.
(For inactivated pixels, $\nu_0=0$ and $\psi(p;\nu_0)$ is the
uniform density. We reiterate that $\psi(p;\nu_1)$ and $\psi(p;
\nu_2)$ are the densities corresponding to the two activated
components in our generative model.) We use $\psi(p;\nu_k)$s rather than
$b(p;\alpha_k,\beta_k)$s in simulating our datasets in order to gauge
performance when the $p$-value mixture distributions 
are only approximately modeled 
by~\eqref{obsllhd}. Therefore, our choice of simulation model does
not match the model on which our estimation methodology and software
are built, but as a result, provides us more real-world performance
assessments. Note also that RF, the FAST methods, pTFCE and spatial
mixture modeling are meant for use 
on Gaussian SPMs, so the generative model may be considered to be
somewhat more favorable for their frameworks.  For these methods which
detect activation using SPMs, we converted each of our simulated
$p$-values to the corresponding upper standard normal quantile before
application. 

We performed experiments in scenarios ranging from easy 
(typical in low-level,
easily-differentiated motor task experiments) to 
difficult activation detection scenarios (as in higher-level 
tasks requiring finer levels of cerebral processing). We obtained
these scenarios by relating the values of $\nu_0\equiv 0$, $\nu_1$ and
$\nu_2$ to the pairwise 
overlap measure $\omega$~\citep{maitraandmelnykov10} calculated
between the unique pairs of densities characterized  by 
$\{\nu_0,\nu_1,\nu_2\}$. Specifically, given $\nu_0=0$ and for each of the
$\bpi$s in Figure~\ref{fig:phantom}, we chose $(\nu_1, \nu_2)$ such that the
overlap between every pair of the three mixture components
was set to be $\omega$ for  $\omega \in \Omega = \{0.01, 0.1, 0.25, 0.5, 
0.75, 0.95\}$. Thus, our simulated mixtures had components ranging from
mildly well-separated ($\omega = 0.01$) to essentially
indistinguishable ($\omega = 0.95$). We call $\omega$ the {\em identification
complexity parameter} and regard it as a surrogate for the difficulty
of the particular activation detection problem, with higher values
reflecting greater difficulty in 
detecting 
activation. 
Figures~\ref{fig:simu_example_0}, \ref{fig:simu_example_1}, and
\ref{fig:simu_example_2} display sample mixture distributions and
simulated statistical parametric maps (SPMs) for three $\omega$s, and
for each of the three phantoms. These SPMs were obtained from our
simulated datasets by using the inverse standard Gaussian cumulative
distribution function on the simulated $p$-values at each voxel. 

\begin{figure}[h]
  \begin{center}
    \includegraphics[width=\columnwidth]{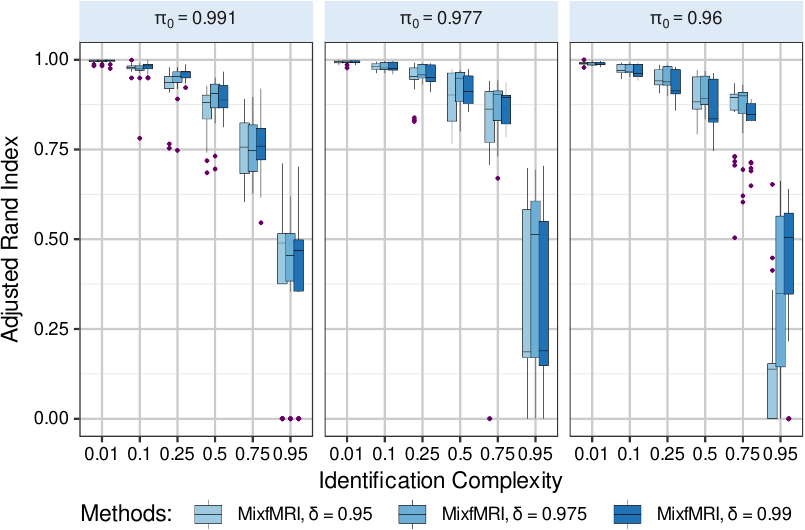}
    \caption{The adjusted Rand index summarizing our method's ability
      to identify the different kinds of activation in each phantom
      for each $\delta$ and $\omega$.} 
    \label{fig:simu_main_adjR}
  \end{center}
\end{figure}

We simulated 25 datasets at each $\omega\in\Omega$ for each phantom, and
obtained activation maps using our APECMa-EGM algorithms and its competitors. 
Recognizing 
that 
a practitioner may not (be able to) specify $\delta$ with complete
accuracy, we evaluated performance of   our method
, and its counterpart that ignored the spatial penalty, with 
$\delta \in \{0.95, 0.975, 0.99\}$ to 
evaluate sensitivity of our results to such potential $\delta$-misspecification.
For all experiments, we set 
$K_{\max} = 8$. 
\paragraph{Simulation results}
We now detail and discuss our results.
\newline
\textbf{\underline{\it Sample illustrations}:}
\begin{figure*}[h]
  \mbox{
    \subfloat[$\pi_0=0.991$]{\includegraphics[width=.33\linewidth]{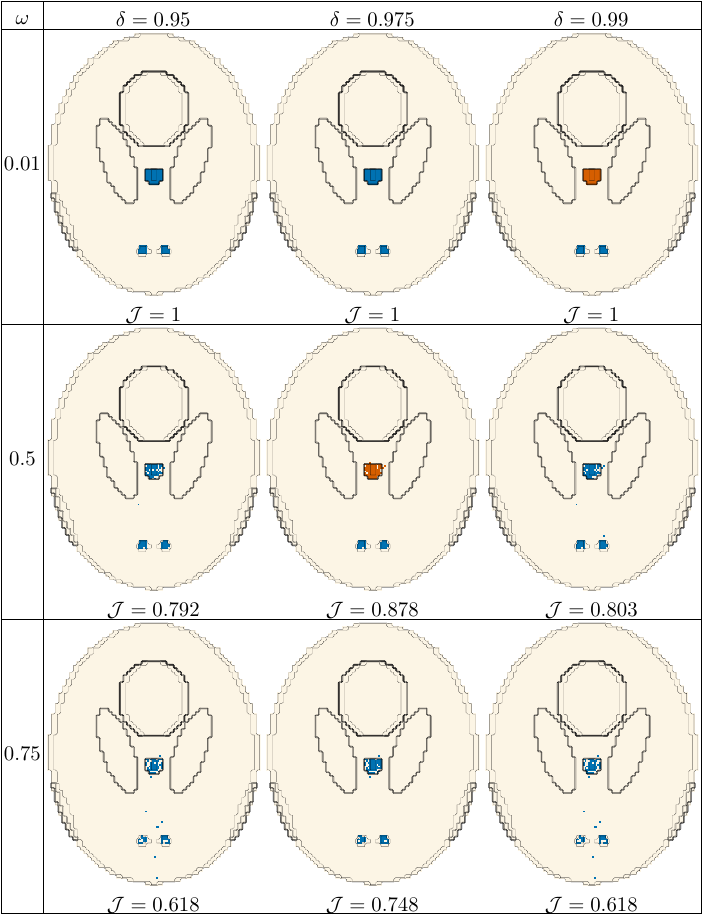}}
    \hspace{-0.0\textwidth}
    \subfloat[$\pi_0=0.977$]{\includegraphics[width=.33\linewidth]{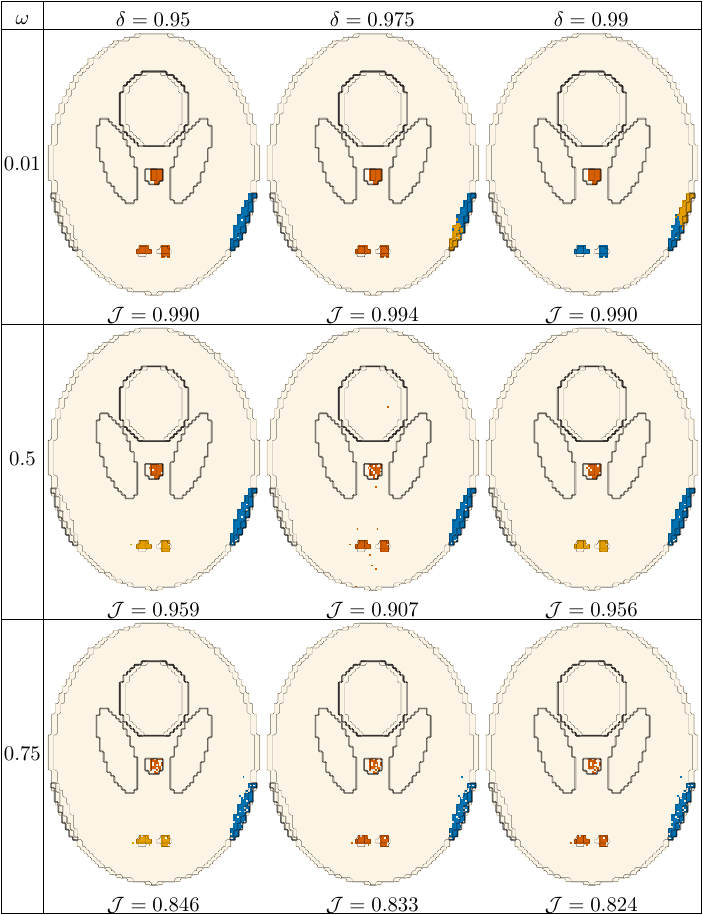}}
    \hspace{-0.0\textwidth}
    \subfloat[$\pi_0=0.96$]{\includegraphics[width=.33\linewidth]{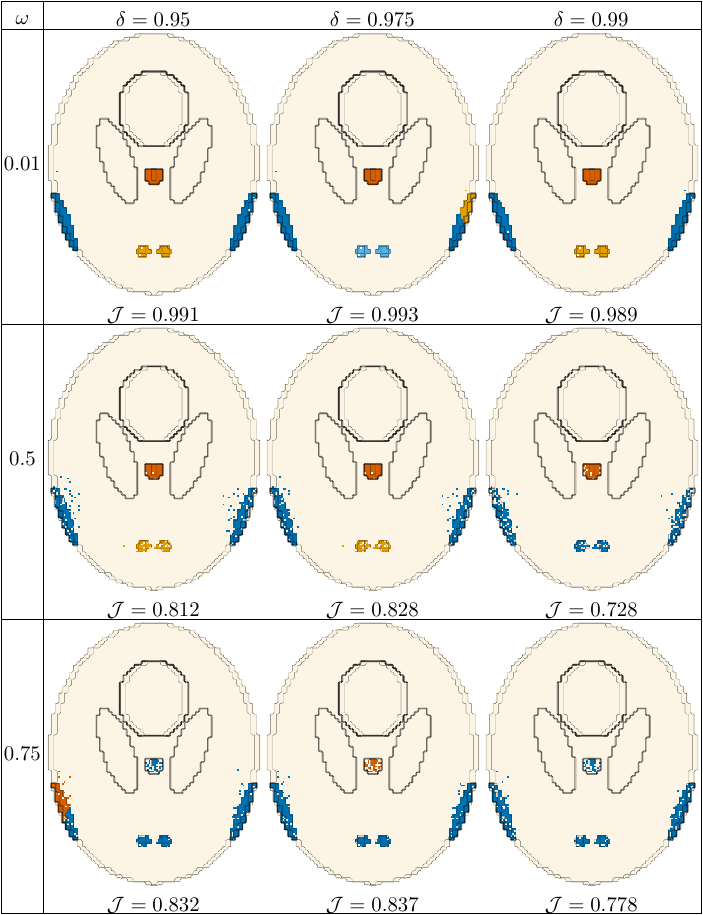}}
  }
  \vspace{-0.1in}
  \caption{Activation obtained using our method on  sample
    realizations obtained using different $\omega$ for the three
    phantoms in Figure~\ref{fig:phantom}.}
    \label{fig:sim}
  \end{figure*}
  Figure~\ref{fig:sim} illustrates activation detected using 
our  method on sample realizations (see
Figures~\ref{fig:simu_example_0}, .~\ref{fig:simu_example_1}
and~\ref{fig:simu_example_2}) of each of the phantoms in
Figure~\ref{fig:phantom}, and with identification complexity
parameter ($\omega$) ranging from the easy ($\omega = 0.01$) to the
substantially ($\omega=0.5$) to very difficult ($\omega = 0.75$).
Figure~\ref{fig:simu_example_0} also has
activation detected using (where applicable, the best-performer among)
the other methods for the simulated dataset corresponding to
Figure~\ref{fig:phantom}a, while 
Figures~\ref{fig:simu_example_1} 
and~\ref{fig:simu_example_2} have the corresponding results for 
Figures~\ref{fig:phantom}b  and ~\ref{fig:phantom}c. (AM-FAST generally does
poorly in many of these cases, so is excluded in these individual
displays for compactness.) Our  approach is generally 
a top performer in each of the examples, and this better performance is
sustained even with increasing activation detection difficulty
($\omega$). However, in these examples, the SMM of~\citet{hartvigandjensen00} 
is often the best performer. Cluster thresholding
(CT-2nd or CT-1st, that is not shown), and FDR also do excellently, 
but under very low $\omega$. With very 
localized true activation (corresponding to the phantom in
Figure~\ref{fig:phantom}a), ALL-FAST does relatively
well even when $\omega$ increases
(Figure~\ref{fig:simu_example_0}). For the other phantoms, pTFCE also  
does relatively well with low $\omega$, but only SMM, ALL-FAST and AR-FAST 
are competitive with higher identification complexity. Despite the good
$\mJ$s, we see that the FAST methods overestimate the spatial extent of
activation. Permutation-testing is a poor
performer across the board, in these examples. In general, many of the alternative  methods degrade 
quicker in performance than our method.  Further, setting 
only the $\delta$ constraint provides good performance, but not
always, and in any case, the improvement is not as much as that
obtained when also including 
the spatial penalty. Finally, the unique ability of our methodology to
potentially detect the different kinds of activation shows some
promise. The results displayed here are on one candidate 
realization at the sample settings, so we now report performance of
all the methods in our large, comprehensive simulation study.
\newline
\textbf{\underline{\bf Comprehensive evaluations}:}
Figure~\ref{fig:simu_main_adjR}  
summarizes the ability of MixfMRI to identify the
different intensities of activation, for $\delta\in\{0.95, 0.975,
0.99\}$. We see that while there is degradation in this ability with
increasing identification complexity, the performance is very good
at lower $\omega$s. Indeed, but for the phantom (with
$\pi_0=0.991$) having very little true activation, the ARI is quite
high even for $\omega$ as high as 0.75. Thus, the performance of
MixfMRI in recovering the different kinds of activation is 
encouraging. Most other methods do not have this ability, mostly providing
only a binary classification of each voxel as activated or not, so we
now evaluate all the methods in terms of $\mJ$. 
\begin{figure*}
    \begin{center}
    \mbox{\subfloat[The Jaccard index summarizing results of all
      methods on all replicates in our 2D simulation
      experiments. The row panel indicates the 
      phantoms (indexed by       the  true $\pi_0$). Each group
      of boxplots corresponds to a given identification complexity
      ($\omega$), with the methods in 
      each group in the same order as the labels in the figure.]{\includegraphics[width=\linewidth]{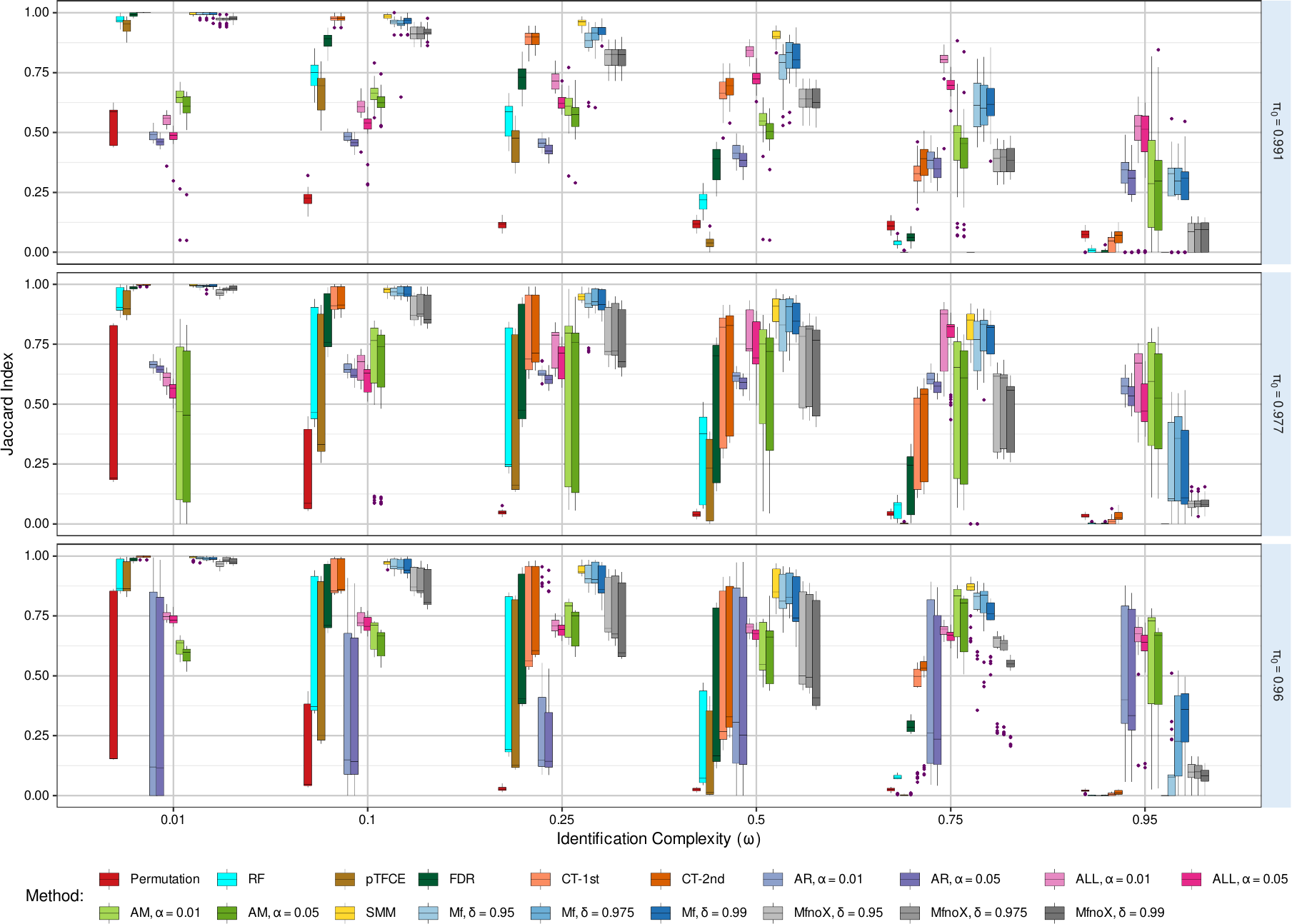}\label{fig:simu_main_JI}}}
    \mbox{
      \subfloat[The $q$-value obtained from the two-sided paired Wilcoxon signed
      rank test of the $\mJ$s obtained using our method (on
      the horizontal axis, for a given $\delta$) and each competitor
      (on the $y$ axis) for each phantom and experimental setting (here
      $\omega_\alpha$ denotes identification complexity of
      $\alpha$). In all cases, darker values indicate 
      significant differences between the competing methods. 
      The yellow-to-green map displays the $q$-value for the cases when our method is better than a given competitor for a specific
      experimental setting, while the
      yellow-to-red map displays the $q$-value for the cases
      where our method is worse. (The $q$-value scale in our display
      is piecewise linear, with a steeper gradient in \mbox{[0,0.05]} than in \mbox{[0.05,1]}).]{\includegraphics[width=\linewidth]{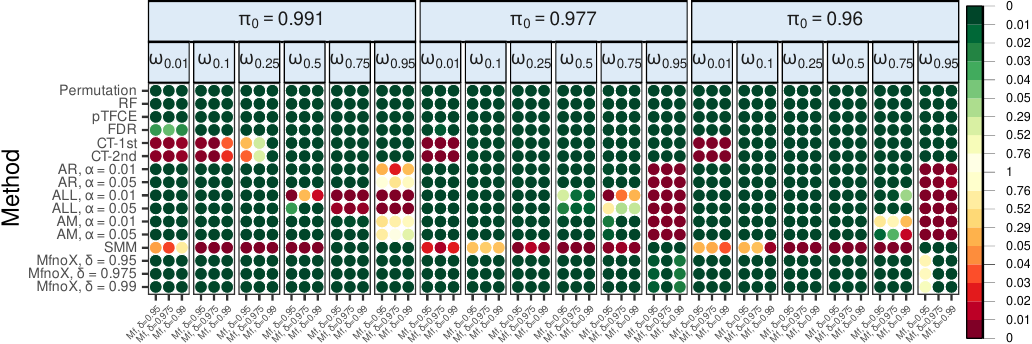}\label{fig:simu_signif_JI}}}
\caption{Comprehensive summary of the results of our 2D simulation
  experiments. The methods are in the same order in both figures and
  use the abbreviations outlined at the beginning of Section~\ref{sec:simulations}.}
\label{fig:simu_main_JIoverall}
\end{center}
\end{figure*}

 Figure~\ref{fig:simu_main_JIoverall} summarizes the results of all the
 two-dimensional simulation experiments, with Figure~\ref{fig:simu_main_JI}
 displaying the performance  of all methods on the three phantoms and
 under different $\omega$s through the
 $\mJ$s. Figure~\ref{fig:simu_signif_JI} provides the
 $q$-value~\citep{benjaminiandhochberg95} of the paired Wilcoxon
 signed rank test for whether the $\mJ$s obtained by MixfMRI
 with a given $\delta$ is different from any of the competing
 methods. (The display differentiates between when MixfMRI is better
 or worse, and shows that we are significantly better (dark greens)
 than the competitor in most cases.)
 For easy  problems ($\omega = 0.01$),
Figure~\ref{fig:simu_main_JI} repeats the trends shown in Figure~\ref{fig:sim}, and does not show much important distinction between our
methodology and the best-performers such as CT-1st or CT-2nd (which report
near-perfect $\mJ$s), or the spatial mixture model,  even though many of these
differences are statistically
significant~(Figure~\ref{fig:simu_signif_JI}). The FAST methods are
generally substantially inferior at these low $\omega$s and continue
to perform indifferently for 
moderate $\omega$, but they, and in particular ALL-FAST, perform very well for
higher $\omega$ and with very small amounts of activation (the phantom of
Figure~\ref{fig:phantom}a). Interestingly, 
\citet{almodovarandmaitra19} suggested using $\alpha=0.01$
and $\alpha=0.05$ for low- and high-noise situations,
but in these experiments, the FAST methods with $\alpha=0.01$
generally had higher $\mJ$s than those obtained using
$\alpha=0.05$.
While all other methods degrade with increasing
identification complexity, our method performs  better relative to
the alternatives. Even with $\omega=0.75$, the median $\mJ$ was
very good for our methods, though there was greater variability in
performance. The spatial mixture model approach was the best performer
in these 2D experiments in all but for the highest $\omega$ (and in
many cases, significantly better than our methods). As in the
case of the single realization case earlier,  permutation testing was
the worst performer across all $\omega$s and phantoms.
 Finally,  including the spatial penalty in
our method improves performance over our method implemented without
the spatial penalty, so for concision and ease of 
presentation, we drop displays and discussion on the case that only
involves the proportion-of-inactive-voxels constraint and not the
spatial penalty.

Our comprehensive evaluations on 2D simulated single-subject
scenarios show competitive performance of our methodology that
incorporates both {\em a   priori} expected proportion of activated
voxels and spatial context. Our methodology is robust and a good
performer even under high identification complexity, which
portends well for its use in experiments involving
high-level cognitive tasks, as in the case of our showcase application.
Additionally, as seen in Figures~\ref{fig:sim}
and~\ref{fig:simu_main_adjR}, we can identify the different
intensities of activation fairly well. We now evaluate performance in
realistic 3D simulation settings. 

\subsubsection{Three-dimensional framework}
\label{sec:neurosim}
\begin{figure}[h]
  \begin{center}
    \includegraphics[width=\columnwidth]{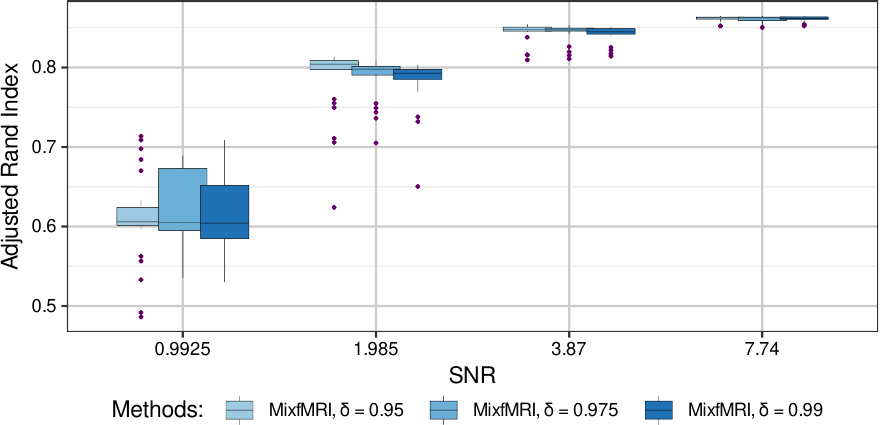}
    \caption{The adjusted Rand index summarizing our method's ability    to identify the different kinds of activation at each SNR, for different $\delta$.}
    \label{fig:neurosim_adjR}
  \end{center}
\end{figure}

Our 3D simulation experiments used
the~\pkg{neuRosim}~\citep{welvaertetal11}~\proglang{R} package that
simulates time series fMRI data according to a desired
experimental paradigm. We used the simulation setup presented
in~\citet{welvaertetal11} that mimics the real-life repetition priming
experiment of \citet{hensonetal02}, with the exception that the radii
of the five activated regions on Page 12 of \citet{welvaertetal11} were each
decremented by one to match 
a more realistic proportion (2.74\%) of in-brain activated voxels~(and
that in the real-life study in \citet{hensonetal02}). We
used four different SNRs of 0.9925, 1.785, 3.87, and 7.74 to 
to characterize varying levels of noise that can impact activation
detection ability. 
Our evaluation was on the SPM formed by the
faces-versus-baseline contrast, with the SPM converted to a $p$-value
map where needed, such as for our methods. For each SNR, we simulated
25 four-dimensional fMRI time 
series datasets, and obtained the corresponding 25 SPMs and $p$-value
maps. Activation was detected using our method (with $K_{\max}=11$)
and its competitors. Figure~\ref{fig:neurosim_adjR} shows that the ARIs are modest
under very low noise, and quite high in all other cases, indicating
good ability of our method in determining the different kinds of
activation.
\begin{figure}
        \vspace{-0.3in}
  \begin{center}
    \mbox{\subfloat[ The Jaccard index  of all the methods evaluated
      on 25 SPMs obtained from fMRI time series data simulated using
      \pkg{neuRosim}, under different SNR. Methods are listed in
      increasing order of their overall median $\mJ$.]{\includegraphics[width=\columnwidth]{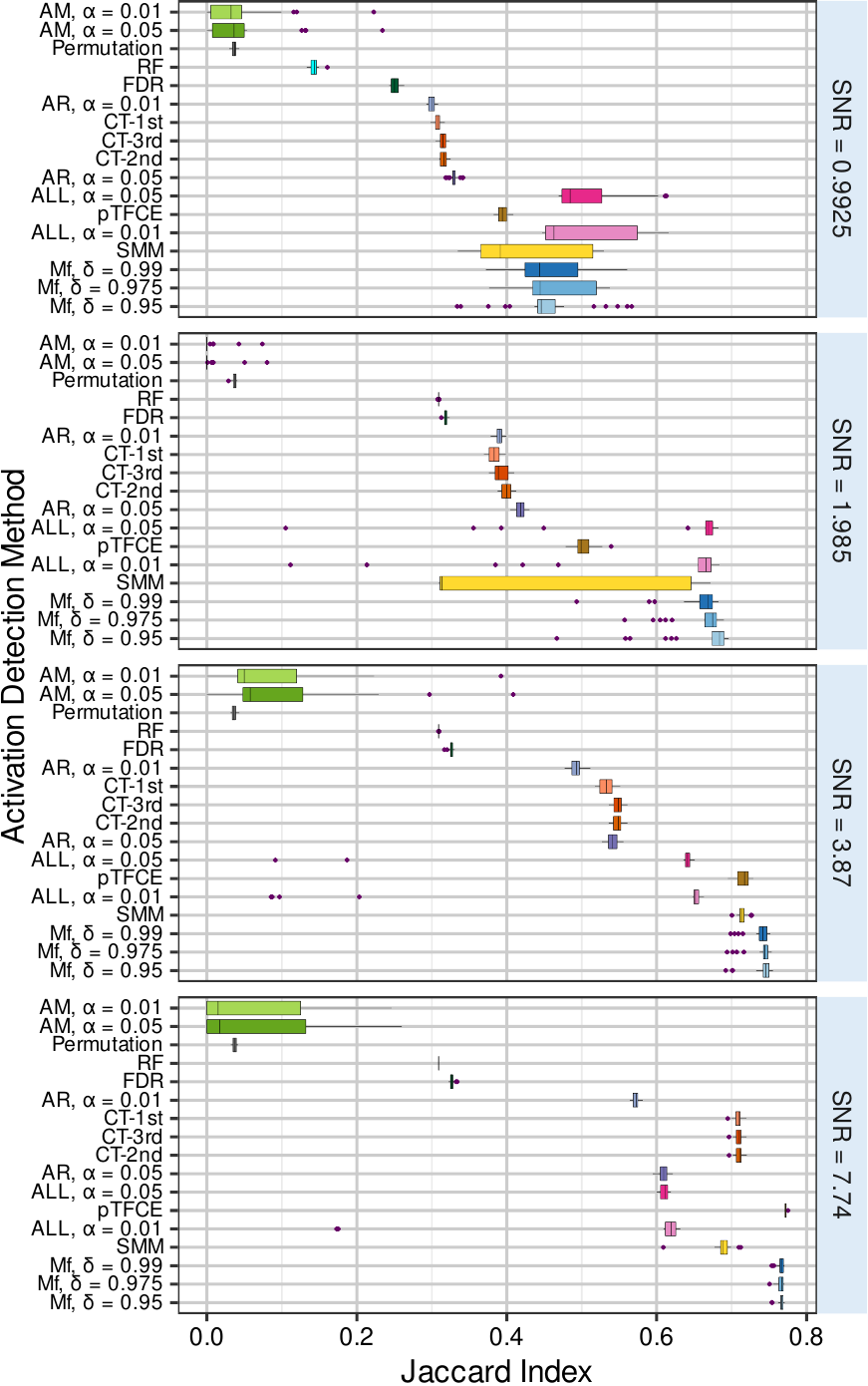}\label{neurosim-JI}}}
    \vspace{-0.2in}
    \mbox{\subfloat[The $q$-value of the two-sided paired Wilcoxon
      signed rank test of the $\mJ$ obtained using our method
      (horizontal axis) at a given $\delta$ and each competitor (in the same order as in Figure~\ref{neurosim-JI}). Displays and scales are as
      in Figure~\ref{fig:simu_signif_JI}.]{\includegraphics[width=0.8\columnwidth]{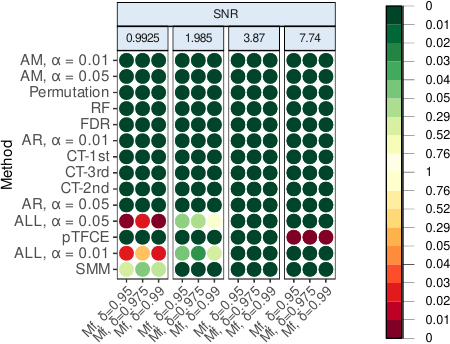}\label{neurosim-signif}}}
\end{center}
\caption{Comprehensive summary of our 3D simulation experiments.}
\label{fig:neurosim}
\end{figure}
Figure~\ref{neurosim-JI} displays the $\mJ$s obtained using our method and
its competitors, with the significance of the differences between our
method and its competitors summarized in
Figure~\ref{neurosim-signif}. The ALL-FAST methods are the best performers
among the competitors (and in a few cases even significantly better
than our methods for the lowest SNR), but not significantly different
(even if slightly worse) than our methods for moderately low
SNR. Also, SMM is worse, but not significantly so, than our method for
very low SNR. Further, while pTFCE is significantly better than our
method at the highest SNR level, Figure~\ref{neurosim-JI} indicates that
this difference in performance may not be significant. Against all other
methods, our method is a substantially better performer, and signficantly so.
AM-FAST and permutation testing are the worst performers at all
SNRs. An interesting observation is the poorer performance of SMM than
in the 2D simulation studies earlier.

Our experiments show good performance of our methodology for a large
range of realistic 2D and 3D simulation experiments. In some cases, we
are bettered by the competitors, with, and especially in 3D settings,
the differences being sometimes significant but not always
important. On the other hand, we are better in most cases, and in many
of these cases the differences are both significant and substantial
enough to be important. Our methodolody is also seen to do fairly well 
in detecting the different kinds of activation.  We now evaluate 
performance of the method and its competitors over some real data examples.

\subsection{Performance on Real Data Examples}
\label{sec:experiment.real}
Our next tranche of evaluations is on 81 datasets
from three sets of fMRI studies. We first evaluated the performance of our
algorithm in null activation 
resting state scenarios for each of 31 typically developing
children~\citep{nebeletal14}. The second set of experiments was from
the right- and left-hand finger-tapping
study~\citep{maitraetal02,maitra09a,maitra10,almodovarandmaitra19} of a
normal male volunteer over 12 sessions spanning two months. 
The final set of experiments were from a Flanker
experiment~\citep{kellyetal08} on 26 right-handed adults.
For  all but the spatial mixture model and the FAST methods, we
spatially smoothed the SPMs obtained in each 
of the experiments  using the automated robust method of
\citet{garcia10} that uses the generalized cross-validation to
adaptively estimate the smoothing parameter. These smoothed maps were
used to obtain the $p$-value maps of activation or significance at
each voxel. We do not apply smoothing to the SPM for the FAST methods
where smoothing is an integral part of the adaptive smoothing and
thresholding algorithm. Similarly, \citet{hartvigandjensen00} also
express concerns about smoothing of the SPM before
activation detection, and so we report on their results obtained
without prior smoothing.  (Each of these methods were also seen to perform
better without prior smoothing of the SPM, while the other methods
performed better with the prior smoothing of the SPM.)   Further, there is no
ground truth available for the finger-tapping and the
Flanker task experiments, so for each experiment, we used the
summarized Jaccard index ($\ddot\mJ$) of 
~\citet{maitra10} to measure consistency in the activation identified
between the different replicates.
The summarized Jaccard index, proposed by \citet{maitra10} for evaluating consistency in several
activation maps, first constructs a matrix $\bOmega$ of the Jaccard
index between every pair of activation maps and then calculates
$\ddot\mJ$ over all the ($M$) activation maps 
in terms of its largest eigenvalue $\omega_{(1)}$. Formally,  $\ddot\mJ =
(\omega_{(1)} - 1)/(M-1)$. It is easy to see~\citep{maitra10} that
$0\leq \ddot\mJ \leq 1$ with $\ddot\mJ =0$ when there is no
agreement at all between any of the maps, because then $\bOmega$ 
is an identity matrix. Further,  $\ddot\mJ=1$ when
all the activation maps are in complete agreement with each other. We
now describe and discuss the results in each experimental study.
\subsubsection{Resting state datasets}
\label{sec:restingstate}
\begin{figure}[h]
  \begin{center}
\mbox{
  \subfloat{\includegraphics[width=\columnwidth]{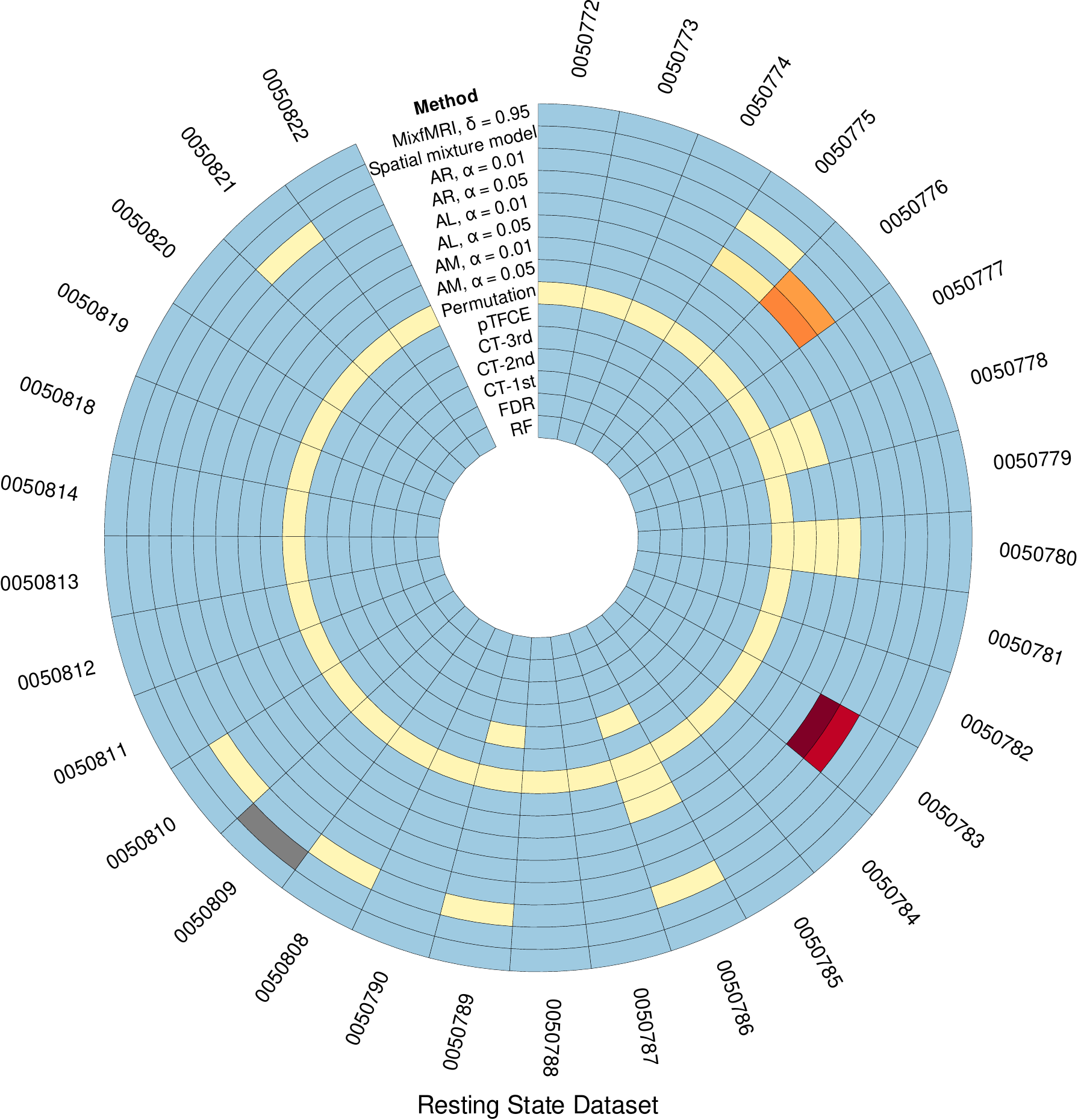}}}
\mbox{
    \subfloat{\includegraphics[width=\columnwidth]{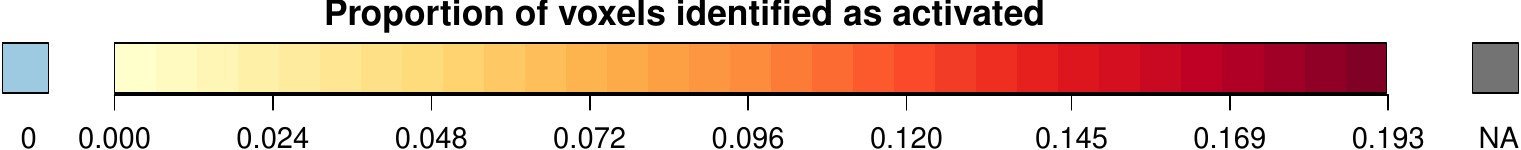}}}
\end{center}
\caption{The proportion of in-brain voxels identified by each of the
  competing methods as activated in  each resting state dataset.
  Each
  dataset is indexed according to its seven-digit identifier. The
  greyed-out cell for the dataset ``005089'' reflects the inability of
  the spatial mixture model to decide on activation for this dataset,
  even after attempts with different parameter combinations.}
\label{fig:RestingState}
\end{figure}

The data used in this set of experiments are from the resting state fMRI
scans of 31 typically developing adolescent children that were acquired using a
single-shot, parallel (SENSE) gradient-recalled echo planar sequence
(TR = 2500 ms, TE = 30 ms, and a flip angle of $70^\circ$) over 156 time points, with the image volumes at each
time-point having axial slices of 3mm each and no slice gap. We used
AFNI~\citep{cox96,coxandhyde97,cox12} to pre-process the dataset and then fit a 
general linear model with autoregressive errors for a purported auditory experiment to the data. The SPMs obtained from each dataset, and their associated
$p$-values were then submitted to our methodology ($K_{\max} = 11$) and
its competitors to obtain activation maps. Figure~\ref{fig:RestingState}
provides a heatmap of the proportion of voxels identified in each
dataset (indexed by its unique seven-digit identifer) as activated
by each method. (For our method, we only report performance with
$\delta =0.95$ since our results obtained using other values of $\delta$
were identical.) No activation was detected, correctly, using our
methodology, RF, pTFCE, cluster 
thresholding, or FDR. The spatial mixture model approach correctly
found no activation in 30 of the datasets, but also was unable to
provide a solution for the remaining dataset. The FAST methods find
varying amounts of activation in at least some of the SPMs, while 
permutation testing detected a very small proportion of active voxels in
every dataset. \citet{eklundetal16} has stressed the importance of
evaluating the false positive rate of activation detection algorithms
on resting state datasets:  we find it encouraging that our method is
among those that passes this test here with each of the 31 SPMs.
\subsubsection{Finger-tapping experiments}
\begin{figure}[h]
  \begin{center}
    \includegraphics[width=\columnwidth]{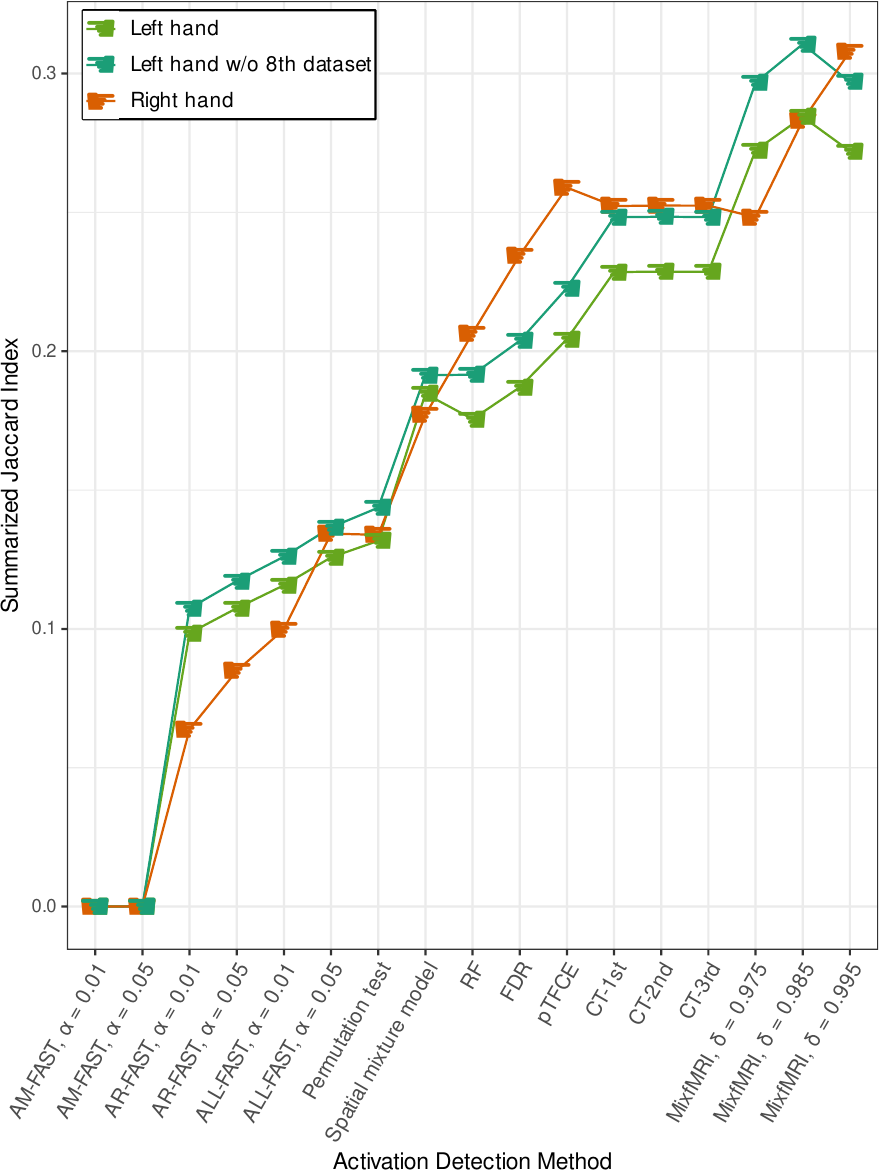}
\end{center}
\caption{ The summarized Jaccard index ($\ddot\mJ$) for assessing the
  similarity in 
  the activation maps obtained by each method for the 12 right- and left-hand
  finger tapping experiments. Because of concerns~\citep{maitra10},
  that the right-hand dominant male may have been
  tapping his right-hand fingers in the eighth dataset, $\ddot\mJ$
  is also obtained for activation maps from the left-hand
  datasets sans the eighth replication.} 
\label{fig:finger}
\end{figure}

The next set of experiments involved the 12 replicated SPMs of
\citet{maitraetal02} from the right-hand and left-hand finger tapping
study of a right-hand-dominant male adult. Figure~\ref{fig:finger} displays
\label{sec:finger}
the $\ddot\mJ$ obtained for the 12 replications on each hand. For this
experiment, we used $K_{\max}=5$ and higher values of $\delta = 0.975,
0.985, 0.995$ to reflect our view that this is a motor experiment with
few and very focused activation areas (and around 1-2\% expected
proportion of activated voxels). For the right-hand experiments, our
method at $\delta = 0.975$  is marginally
worse ($\ddot\mJ=0.248$) than that obtained using CT ($\ddot\mJ=0.252$) or 
pTFCE ($\ddot\mJ=0.259$), however, our 
results obtained using $\delta=0.985$ ($\ddot\mJ=0.283$) and 
$\delta=0.995$ ($\ddot\mJ=0.308$) are substantially better than all 
competitors. For the left 
hand, we see that our method outperforms all its
competitors. (\citet{maitra10} has expressed some concern about the quality
of the data in the eighth left-hand experiment replicate, therefore we also
evaluated $\ddot\mJ$ of the results after excluding this
experiment. The $\ddot\mJ$ for almost all methods
improve, but even here, MixfMRI is by far the best performer.) For
these datasets, cluster thresholding does quite well, and is the
next-best performer (after MixfMRI) in the left-hand experiments. The
spatial mixture model approach performs modestly on these experiments,
and is mildly to substantially (for the right-hand experiments) worse
than FDR. AM-FAST, AR-FAST, ALL-FAST and permutation testing are, in
that order, among the worse performers. An encouraging aspect of the
$\ddot\mJ$s for our method is its consistency in the right hand, 
that indicates that a higher restriction on $\delta$ identifies more
similar areas of activation in each map.
For the left-hand experiments, $\ddot\mJ$ goes up from $\delta =
0.975$ to $\delta=0.985$, but dips slightly from there to
$\delta=0.995$, potentially attributed to the fact that this is a
left-hand finger-tapping experiment performed by a right-hand dominant
male. However, whether with the right or the left hand, our
methodology is overall the better-performing method. We also comment
that the generally low values of $\ddot\mJ$ for all methods provide
support to concerns~\citep{maitra10,almodovarandmaitra19} on
potential issues in quality and preprocessing of this dataset.
\subsubsection{Flanker task experiments}
\label{sec:flanker}
\begin{figure}[h]
  \begin{center}
    \includegraphics[width=\columnwidth]{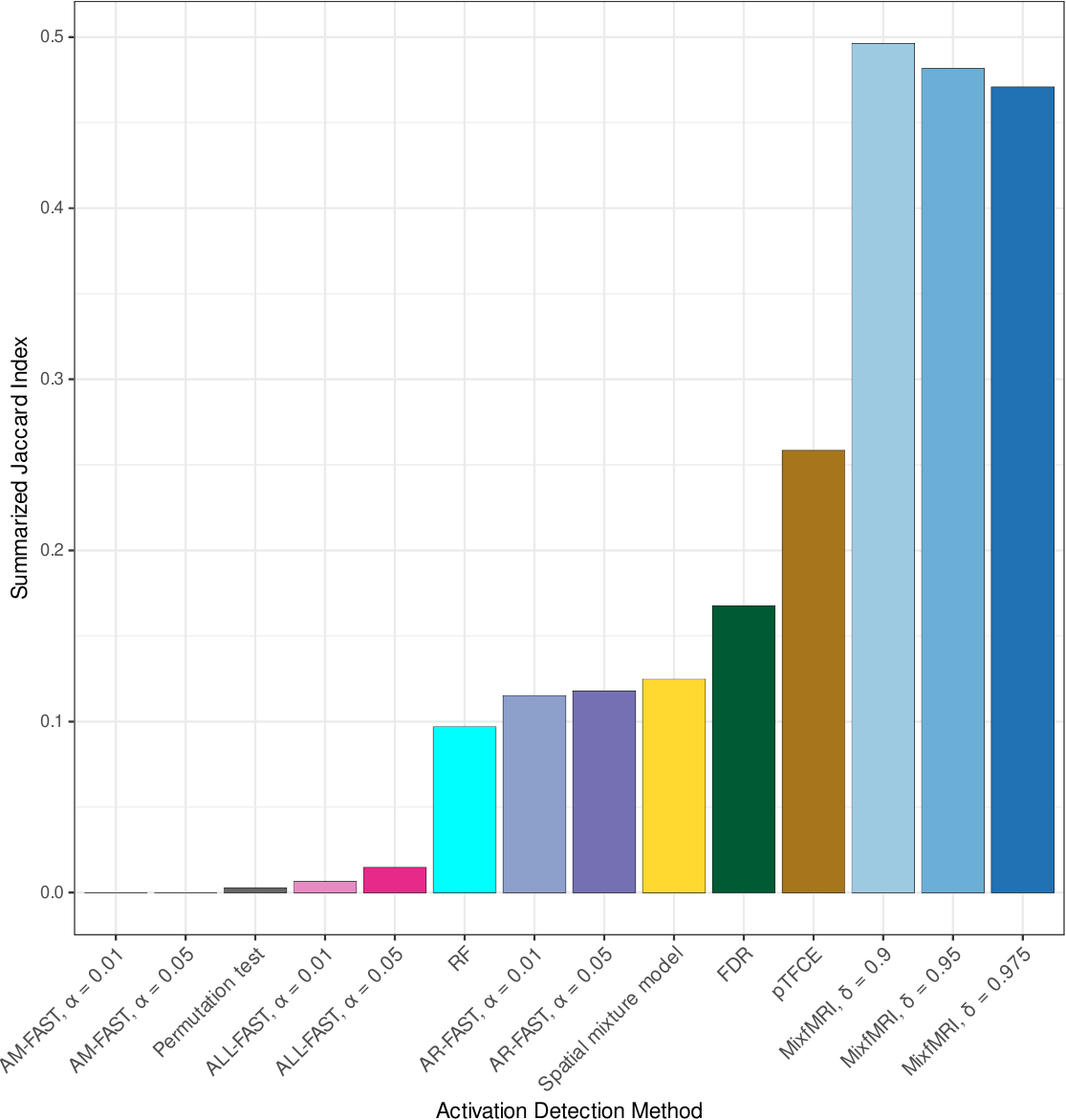}
\end{center}
\caption{The summarized Jaccard index for assessing the similarity in
  the significance maps obtained by each method for the 26 Flanker
  task experiments.}
\label{fig:flanker}
\end{figure}

The Flanker task data are from a study~\citep{kellyetal08} on 26
right-handed adults who were vicenarians or tricenarians, and were
imaged using echoplanar 
imaging (TR=2000 ms, TE = 30 ms, flip angle = 80$^\circ$) over 146
time-points to get, 146 $64\times64\times40$ image volumes of voxel size
$3\times3\times4\mbox{mm}^3$, while
participating in 12 congruent and 12 incongruent Eriksen flanker task
trials, that were presented to them at varying
intervals of 8 to 14 seconds, in pseudo-random order. We used AFNI to
first preprocess the dataset to a common MNI template for each subject, and
then fit a 
general linear model with the appropriate design matrix and first-order autoregressive moving average
errors. From each dataset, we obtained a SPM that captured the
contrast between the congruent and the incongruent task, and obtained
the corresponding $p$-values at each voxel. This is an example of an
experiment with a two-sided alternative hypothesis, and our method is
applied using the additional development of
Section~\ref{sec:2sided}. 
Once again, there is no ground truth for comparison, but we note that
although these SPMs are for different subjects, they are all of normal
younger adults and in standardized space, and so we evaluate
performance by calculating $\ddot\mJ$ between the activation (or
significance) maps obtained on each SPM by each method. Here, the
activation map specifies the significance or otherwise of a contrast,
as determined by the applied method. (Here, we evaluate \pkg{MixfMRI}
with $K_{\max}=11$, and  using $\delta=0.9, 0.95, 0.975$, with the
higher values reflecting the circumstance that this is a two-sided
hypothesis. Also, memory requirements in the \pkg{AnalyzeFMRI}
implementation of cluster thresholding precluded us from evaluating any of the
CT-1st, CT-2nd, or CT-3rd methods so we do not include these
methods in our comparisons.) Figure~\ref{fig:flanker}
summarizes the results and shows that our methods yield
$\ddot\mJ$-values that are almost twice as good as its closest
competitor (pTFCE). The performance of FDR,
spatial mixture model and the AR-FAST approaches is middling, while
permutation testing and the other FAST methods  perform poorly. It is
unclear if the very marginal decrease in the 
consistency of the significance maps with increasing $\delta$ are
themselves significant, but if so, may point to the difficulty of
consistency identifying significance in higher-level tasks. 
However, we see that we are by far the best performer in this
two-sided hypothesis testing scenario.  

\subsection{Summary of Results}
\label{sec:expt.summary}
The results of our experiments on both simulation and real data
examples show good performance of our MixfMRI methodology developed in
this paper. In both 2D and 3D simulation setups, our method is always a
top performer. The superlative performance of the spatial mixture
model in the 2D simulation experiments (for all but the highest model
complexity) does not quite translate to the
3D simulation setup, and even more so, to the real-data 3D examples,
where we are almost always the best method. Our method is also among those that
correctly identify no activation in all 31 resting state datasets. Among
the other methods, the second-best performer is unclear in the 3D
simulation experiments, but is pTFCE in the case of the finger-tapping
and Flanker task real-life data experiments. Interestingly, and in general,
the performance of permutation testing for detecting activation was
anemic at best  -- a finding also reported in~\citet{almodovarandmaitra19}
and somewhat contrary to that of~\citet{eklundetal15}.

We close this section with a brief discussion on the specification
of $\delta$, the constraint on which is an integral part and major contribution
of our work.  
Our simulation studies indicate that the choice of $\delta$ slightly
affects performance for all three 2D phantoms and
identification complexities, with greater relative (albeit  marginal)
inaccuracy  when $\delta$ is misspecified downwards ({\em i.e.}
$\delta$ is less than the true $\pi_0$) than with  $\delta$ is
specified to be greater than the true $\pi_0$, where it
still provides quite accurate identification. Given that it is
impractical to expect a practitioner to precisely specify $\delta$, it
is encouraging to note the robustness of our method to the exact
choice of $\delta$.
\section{Activation in sports imagination experiment with implications in treatment and therapy of PVS patients}
\label{sec:applications}
We now re-analyze the sports imagination
experiment dataset of Section~\ref{sec:data}. Recall that the value of
this experiment is in providing for a way to communicate with, and
provide therapy and treatment for PVS patients~\citep{owenetal06}, who
because of TBI or 
other reasons, may have individualized brain structure and function
that needs to be accurately identified. We therefore evaluate
whether we can, as a follow-up to our very encouraging 
results of Section~\ref{sec:simulations}, accurately recover
activation in a known normal subject where the results can be
explained and interpreted against known cerebral structure and
function, and where Section~\ref{sec:data} has suggested unclear
results with current standard analysis methods. We revisit this
dataset by applying our refined methodology to the voxel-wise
$p$-values obtained, as described in~\ref{sec:preprocess}. Here the most 
plausible value of $\delta$ is 0.99 but we also try $\delta = 0.975$
and 0.999  to check 
the effect of $\delta$-misspecification. With $K_{\max}=20$,
BIC (and ICL-BIC) chose $\hat K = 11, 
12$ and 6 for solutions using $\delta =
0.975, 0.99$ and $0.995$. 
In either case, the inactive component did not merge with any of the
other groups in the procedure outlined in the first part of
Section~\ref{sec:merge}. Using the second part of
Section~\ref{sec:merge}, the activated components merged to form four,
five and four clusters with $\delta=0.975, 0.99, 0.995$, respectively. 
We now discuss and interpret the results.
\subsection{Analysis and interpretation of results}
\begin{figure}[h]
  \centering
\subfloat[]{
\raisebox{-0.41\height}{\hspace{-0.1in}\includegraphics[width=\linewidth]{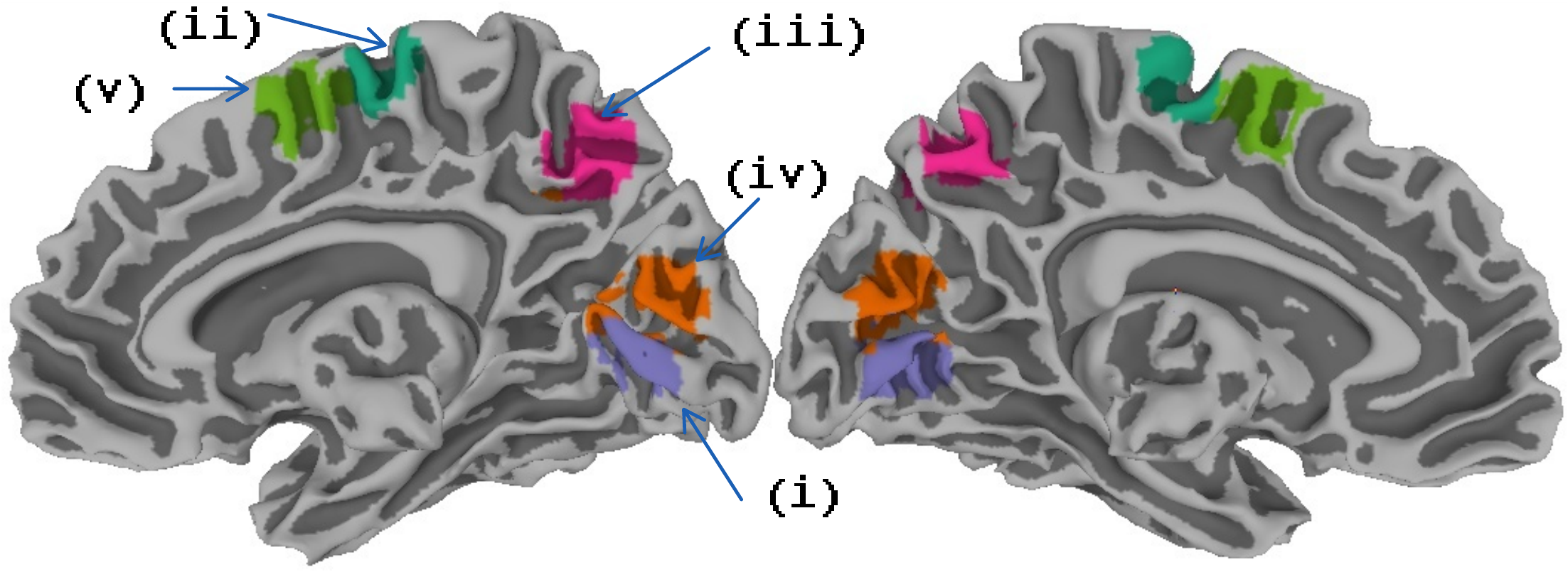}}
}
\hspace{-0.2in}
\subfloat[]{\footnotesize
\begin{tabular}{clrrrr} \hline
Region&Indicator color &  $\bar t_{\mbox{stat}}$ &{\em \#v} &$\hat\alpha$ & $\hat\beta$
  \\ \hline
(i) & Purple   &  3.736 & 178 &0.524 & 404.09\\
(ii) & Sea green  &  3.420 & 417 &0.551 & 140.48 \\
(iii) &Pink   & 3.025 & 387 & 0.695 & 112.27\\
(iv)&Orange   &  2.761 &533 & 1.652 & 181.74\\
(v) &Green   & 2.463 & 175 & 1.533 &  79.55  \\ \hline
\end{tabular}}
\caption{ (a) Activation detected using our method with
    $\delta=0.99$ in a healthy female volunteer performing a sports
    imagination experiment. (b) The five activated regions, 
    identified by their indicator color, 
    the number of voxels ({\em\#v}), 
  the average $t$ statistic 
  and estimated beta distribution parameters for each region. In that
  order, the five regions span (i) the calcarine sulcus and the
  inferior occipital cortex/lingual gyrus, (ii) the PMC, (iii)  the
  pre-cuneus, (iv) the cuneus, and  (v) the SMA. }
\label{fig:imag2}
\end{figure}

We focus our discussion on results obtained using $\delta = 0.99$. 
Figure~\ref{fig:imag2} summarizes the
results and indicates the five distinct regions identified as
activated. We discuss these regions
numbered (i) through (v), in
decreasing order, of their averaged $t$-statistic (highlighted in Figure~\ref{fig:imag2}), recalling that the
experimental paradigm is of a healthy female volunteer imaged
alternately while at rest and while imagining playing tennis.

The region with the highest average (standardized) intensity of activation 
is among the smallest, and spans the calcarine sulcus and the
inferior occipital cortex/lingual gyrus. The primary visual cortex --
where dynamic and static information 
from the visual hemifields is organized coherently for later
processing in the cuneus -- is primarily located in the calcarine sulcus,
while the inferior occipital/lingual gyrus is important for visual
memory, attention, and in orienting the brain to visual location of an
existing or remembered scene.  That these regions would be activated
with high intensity makes sense given that the subject is tasked with
imagining playing tennis, and needs to recall visual imagery of the
activity and in processing information in pretend-planning and executing her
moves. But for the PVC which is identified rather faintly, these
regions are not identified using cluster-wise thresholding, pTFCE~(see
Figure~\ref{fig:imag1}a, b) or many other methods.


The next highest average $t$-statistic is a large region in 
the primary motor cortex (PMC) 
that executes functions planned by the pre-motor areas. This
area also includes a somatotopic map comparable to the somatosensory
cortex, and has been associated with mental motor imagery and
sensorimotor planning in multiple fMRI studies involving imagined
movements~\citep{rothetal96,lotzeetal99,stippichetal02}. Once again,
the fact of these regions being activated makes sense in the context
of the experimental paradigm, however, but for the
tiny blip of activation in the somatosensory cortex, these regions
were largely missed in cluster-wise thresholding or AR-FAST, and completely
by  methods other than pTFCE.

The region of activation with the third highest averaged
$t$-statistic is in the precuneus which is an area of the brain that
is known to be  involved in the  storage of  long-term memory and in
relating current and past experiences,  self-reflection and
cogitation.
Given that the subject is tasked with imagining playing tennis that
has to have her recollect and draw from her past experiences, it
makes sense for the precuneus to show activation. The precuneus was
also identified in the cluster-thresholding of Figure~\ref{fig:imag1}
but the spatial extent was low. This region was also almost missed by
pTFCE. 

The fourth region is the largest activated region and spans the cuneus
that is superjacent     to the calcarine sulcus. This area                                 is  known to pre-process 3D space, color
 and other information while the next region includes the SMA that
 works in concert with the pre-SMA, the striatum, and other regions 
 to collect prefrontally driven information about what the subject is
 doing  and guides appropriate movement 
 in space. This higher-order
information is then sent to the PMC, which eventually
directs  motor movements. Once again, activation of these areas are in
keeping with the experimental paradigm. Cluster-wise thresholding and
pTFCE also identify this region, though the spatial extent of these
regions is lower. 

The fifth region is the smallest activated region and is mostly the
Supplementary Motor Area (SMA). This region works in concert with
pre-SMA, the striatum, and other regions to collect prefrontally
driven information about what the organism is doing and where it
should move in space to execute its goals. This higher order
information is then sent to primary motor cortex, which eventually
send movement commands to the limbs, face, and trunk. Once again, it
is reasonable to expect activation in this region, considering the experimental
paradigm, but AR-FAST, cluster thresholding and pTFCE miss this region entirely. 

\subsubsection{Activation detection in sports imagination experiment with
  $\delta = 0.975$ and $\delta = 0.995$}
\label{sec:actmore}
\begin{figure}[h]
  \mbox{
\subfloat[]{
  \includegraphics[width=.5\textwidth]{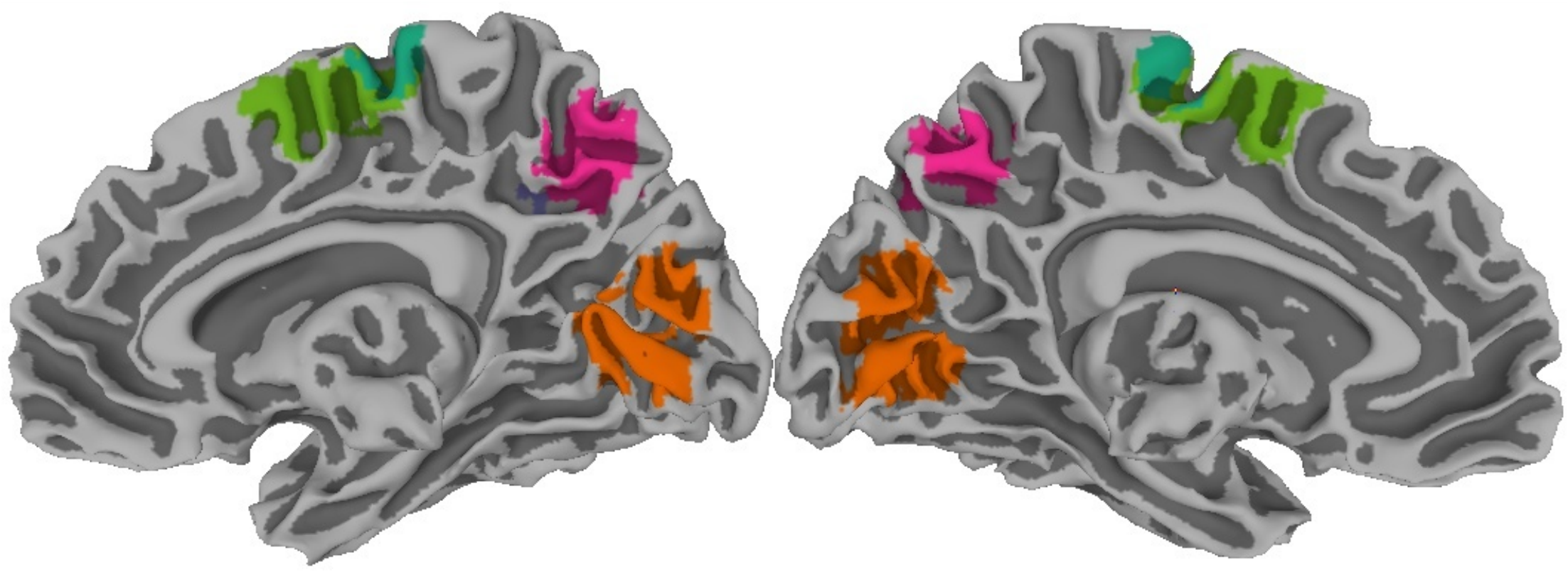}}}
\mbox{
\subfloat[]{
  \includegraphics[width=.5\textwidth]{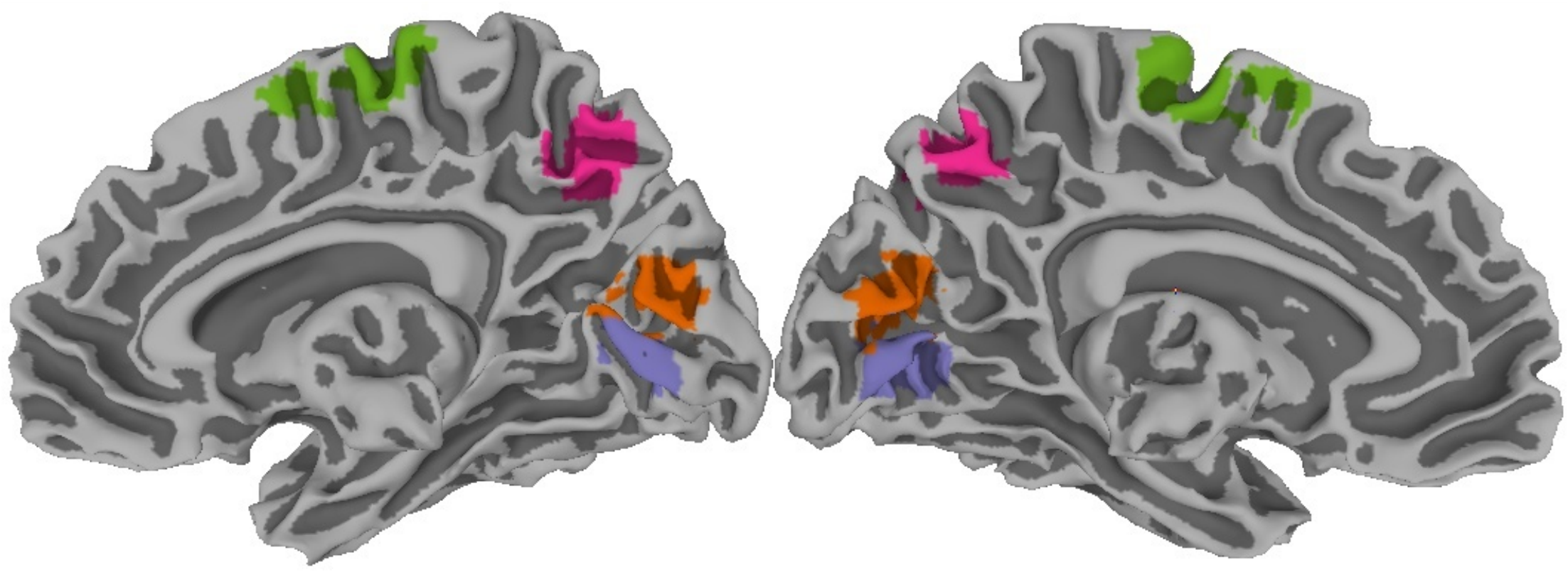}
}
}
\vspace{-0.1in}
\caption{Activation detected in a healthy female volunteer using our
  method with (a) $\delta = 0.975$ and (b) $\delta = 0.995$.}
\label{fig:actmore}
\end{figure}
Figure~\ref{fig:actmore} displays activation regions detected by our
method with $\delta=0.975$ and 0.995. Compared to
Figure~\ref{fig:imag2}, both choices of $\delta$ result in one less
region, but for different reasons. With $\delta=0.975$,
Figure \ref{fig:actmore}a shows one less region than Fig~\ref{fig:imag2},
merging the cuneus with the calcarine sulcus and the inferior
occipital cortex/lingual gyrus, while Figure~\ref{fig:actmore}b has
the SMA and PMC placed in the same activated group. So the results are not
very dissimilar and exhibit robustness with regard to modest
misspecification of $\delta$. 
\subsection{Significance of our findings}
Our results with this imagination dataset indicate that our refinements
that account for both 
spatial context and {\em a priori}  constraints on the expected
proportion of activated voxels can provide improved and interpretable
activation detection in a single-subject sports imagination
experiment. Specifically, when compared to the results from
cluster-wise thresholding, pFTCE or AR-FAST (Figure~\ref{fig:imag1}) that only
identified the first three of our 
regions and did not identify the cuneus which preprocesses information on 3D
space and color, or the SMA that gathers prefrontal 
 information to send to the PMC, our identified activation
 regions are clear and interpretable in the context of the 
experimental paradigm, providing confidence in our findings even
though they are derived from a single subject.
This demonstrated ability of our approach to identify these regions, each
with  different kinds of activation, is a major strength relative to
all existing methods. This subject was a
normal healthy female and so while her activation maps are
interpretable in the context of the study, we would not {\em a priori}
know this, for example with a TBI survivor or a patient in PVS, or more
generally in a  clinical setting.
The original goal~\citep{owenetal06} of sports
imagination experiments in patients in PVS was to facilitate therapy
and provide ways of communication with them but, as explained in
Section~\ref{sec:data}, this 
benefit has not been realized because of murky activation detection
even in group PVS patient experiments~\citep{bardinetal11}. It is essential to
have reliable activation detection at the individual subject level in
order to permit therapy and communication with PVS patients and the
development and analysis of this paper show promise in making this
goal possible. More generally also, our refinements and improvements
can improve activation detection at the individual level, making
it possible to adopt fMRI in a clinical setting.

\section{Discussion}
\label{sec:discussion}
We provide an improved activation detection approach in task-based
fMRI that incorporates spatial context and also the proportion of
voxels {\em a priori} 
expected to be activated as per the experimental paradigm. Our
statistically sound model-based methodology is computationally 
practical and implemented in our \pkg{MixfMRI} R package.
Simulation experiments on settings of low to very high detection
difficulty indicated our methods to have superior 
performance over existing ones. Our methodology also correctly
predicted no activation in null activation settings and was a top
performer, in terms of consistency, or real-world datasets. 
When applied to a single-subject sports imagination experiment study,
with deep implications in the treatment of PVS patients, the method provided interpretable results and improved
activation detection, identifying regions that had not been activated
by current methods. An additional benefit of our approach is the
ability to differentiate intensities and kinds of activation. 
This augurs well for applying our methodology
to clinical settings, such as with TBI survivors or subjects with
impacted or individualized brains where accurate activation detection
is essential to identify pathologies, treatments and therapy.

Our refinements are easily extended to  situations beyond voxel-wise
$p$-values of the SPM. For instance, our methodology could be easily
extended to the SPM itself. Further, even in the context of
$p$-values, we have developed our methodology using an approximate
mixtures-of-beta mode, but we could easily incorporate more precise
models, as for instance provided in~\citet{maitra09b}. Also, our
methodology is really intended for single-subject studies, 
however, it can be easily adapted to apply to activation detection in
multi-subject group studies. In this case, we simply apply our
methodology to the SPM obtained from the
group study.
 Activation detection could potentially be further improved by
semi-supervised approaches that include known 
areas of inactivation, voxel-wise {\em a priori} activation
probability maps drawn from normal subjects, or better processing such
as including phase in the fMRI time-series data analysis~\citep{adrianetal18}. 
Further, it may we worth investigating if the performance of competing methods,
{\em e.g.}, the spatial mixture model, can be further improved by incorporating
{\em a priori} constraints on the expected proportion of activated
voxels. Thus, we see that while we have provided a substantial
contribution to  accurate activation detection in fMRI, a number of
interesting extensions and investigations remain that may benefit from further
attention. 

\section*{Acknowledgements}
The authors are very grateful to two anonymous reviewers 
whose comments on an earlier version of this manuscript greatly
improved its content.  Out sincere thanks also to B. Klinedinst and
A. Willette of the 
Program in Neuroscience and the Department of Food Sciences and Human
Nutrition at Iowa State University for help with the interpretation of
Figure~\ref{fig:imag2}. 
\section*{Funding}
R. Maitra was supported in part by the National      Institute of
Biomedical Imaging and Bioengineering (NIBIB)  of the National
Institutes of Health (NIH) under Grant Nos. R21EB016212 and
R21EB034184. The content of this paper is however solely the
responsibility of the authors and does not represent the official
views of the FDA, the NIBIB or the NIH.

\section*{Data Availability}
The datasets used in the simulations is publicly available within the
authors' R package \pkg{MixfMRI} implementing the methods. The sports
imagination experiment dataset is publicly available at
\url{https://www.jstatsoft.org/index.php/jss/article/downloadSuppFile/v044i11/v44i11-data.zip}. The
processed SPM of $p$-values is available as the {\bf pstats} object in
the R package  \pkg{MixfMRI}.

\section*{Availability of computer code}
The methodology developed in this article is implemented in publicly available
R package \pkg{MixfMRI} publicly available as a ``Contributed package'' at
\url{https://www.r-project.org}. The latest developmental version of
this package is available at \url{https://github.com/snoweye/MixfMRI}.

\appendices
\section{Sports imagination experiment dataset: Additional details}
\label{sec:data.supp}
\subsection{Background}
\label{sec:data.bg}
The time-course sequence of images were
acquired on a MR scanner using a gradient echo-echo planar imaging
sequence with an echo time (TE) of 40 milliseconds, relaxation
time (TR) of 2 seconds and an excitation flip angle of 80 degrees.  At
each time-point, the scanner 
acquired 30 axial slices of 4 mm thickness each and a field-of-view of 24 cm
with an imaging grid of size $64 \times 64$, yielding in-plane voxel
dimensions of 3.75 mm. \citet{bardinetal11} do not provide detailed
discussion and specific analysis of this dataset, but it was later
used (without pre-processing) and publicly 
released by \citet{tabelowandpolzehl11} to demonstrate features of
their \pkg{fmri} software package in \proglang{R}~\citep{Rcore}.
\subsection{Data Preprocessing}
\label{sec:preprocess}
We used the umbrella {\tt uber\_subject.py} in the  Analysis of Functional
Neuroimaging (AFNI) software~\citep{cox96,coxandhyde97,cox12} to 
pre-process, register~\citep{coxandjesmanowicz99,saadetal09} and analyze the
dataset. The script consists of two main steps. First, the
functional image data cube at all time points is registered with the
anatomical data cube. Our registered and aligned 3D dataset had
$46\times 55\times 43$ voxels, of which 33,753 voxels were inside the
brain. There were 105 time-points with TR=2s, and the time series at
each voxel was fit  using the model 
\begin{equation}
Y_t = \beta_0 + \beta_1 s_t + \beta_2 d_t + \beta_3 d_t^2 + \epsilon_t
\label{eqn:arma_rc}
\end{equation}
where, at the $t$th time-point,  $Y_t$ is the observed BOLD response,
$s_t$  is the expected 
BOLD response obtained by convolving the input stimulus and the
Hemodynamic Response Function (HRF),
and $d_t$ is the time drift parameter (that is assumed to have a
quadratic effect on the observed BOLD response), and $\epsilon_t$
is the noise or error in the observation with an assumed 
autoregressive moving average ARMA(1,1) dependence structure. (Here, we have
temporarily dropped the subscript for the voxel for readability but it
is present in all the parameters and the random quantities in
\eqref{eqn:arma_rc}.)
Our voxel-specific parameters allow us to account for local
inhomogeneities -- the voxel-wise fitting of the model is also
computationally practical. At this stage, we do not account for any
spatial structure in the model. Although all parameters are fit, our
interest in this experiment is only in knowing if the parameter $\beta_1$
is positive or not. Framed in terms of a hypothesis testing problem,
we test $H_0: \beta_1 = 0$ v.s. $H_a: \beta_1 > 0$ at each voxel which
means that there is a positive association between the observed
BOLD response at that voxel and the expected BOLD response after
accounting for other factors. All other parameters are nuisance
parameters and of no primary importance.
At each voxel, we test
$H_0: \beta_1 = 0$ against $H_a: \beta_1 > 0$ which means that there
is a significant voxel response to
stimulus when $H_0$ is rejected. Thus, the $t$-statistics for testing
the hypothesis of no activation against the alternative of positive
association with the expected BOLD response are obtained at each
voxel. 

Because our analysis outlined above does not account for spatial
context, the $t$-statistics were 
spatially smoothed for consistency using the automated robust method
of \citet{garcia10} which uses generalized cross-validation to
adaptively estimate the smoothing parameter. The $p$-value of
activation at each voxel was computed at each voxel from these test
statistics. These obtained $p$-values were used to construct
activation maps.

\subsection{Constructing Activation Maps}
The $p$-values at each voxel were thresholded to construct activation
maps using some common  approaches. These maps
and the test-statistics at those voxels were then  
mapped using the SUrface MApping 
tool (SUMA) of ~\citet{saadandreynolds12}. We used FDR with 
$q=0.05$~\citep{genoveseetal02,benjaminiandyekutieli01} and found no
activation, highlighting the challenges faced by
a global thresholding method such as controlling FDR for constructing
activation maps for experiments such as this one.
We also used cluster-wise thresholding methods to detect
activation with AFNI's \proglang{3dClustSim} package to determine the
minimum number of contiguous voxels in each activation region
at a threshold of 0.001 (following \citet{wooetal14})  using first-,
second- and third-order neighborhoods.
\section{Statistical Methodology: Additional Details}
\label{sec:methodology.supp}
\subsection{Parameter Estimation via The EM Algorithm}
\label{sec:parameter_estimation}

\subsubsection{Proof of Result~\ref{theo:result}}
  \label{proof}
Define the  Lagrangian $\mathcal L(\bpi,\lambda_e,\lambda_\iota ) =
g(\bpi;\bW)  - 
\lambda_eh_e(\bpi) - \lambda_\iota h_\iota(\bpi)$ with convex set
$\bPi$. Then $\bpi^\bullet$ is a 
local maximum 
if and only if there exists exactly one $\lambda_\iota ^\bullet$
satisfying stationarity~({\em i.e.} having
$\nabla_{\bpi} \mathcal
  L(\bpi^\bullet,\lambda_e^\bullet,\lambda_\iota ^\bullet) = 0$) and the 
  Karush-Kuhn-Tucker (KKT) conditions~\citep
  {boydandvandenberghe04}. Note that the constraint functions
  $h_e(\bpi)$ and $h_\iota(\bpi)$  are both affine, so all KKT
  conditions are satisfied.
  
The stationarity condition for maximizing $g(\bpi;\bW)$
yields  $\bpi^\bullet$ satisfying the equations:
\begin{equation}
  -\frac1{\pi^\bullet_k}\sum_{i=1}^nw_{ik}
  +\lambda^\bullet_e+\lambda^\bullet_\iota1_{(k=0)} = 0,\qquad
  \mbox{ for } k=0,1,\ldots, K.
  \label{eqn:stat}
\end{equation}
The inequality constraint is inactive  if $\lambda^\bullet_\iota =
0$. This means that the local minimizer  $\bpi^\ast$ of 
$g(\bpi;\bW)$ under the equality constraint ($h_e(\bpi) = 0$) is a
feasible solution for our problem. We now consider the case
of the active inequality constraint, that is, $\lambda_\iota^\bullet >
0$, for which we use the KKT conditions. Primal
feasibility is itself defined by the constraint in (b).  The dual feasibility condition stipulates that 
$\lambda_\iota ^\bullet\geq 0$ so with an active inequality
constraint it is enough to focus on $\lambda_\iota^\bullet > 0$. Combining
this condition ($\lambda_\iota^\bullet > 0$) with complementary slackness
($\lambda_\iota^\bullet h_\iota(\bpi^\bullet) > 0$) yields $\pi^\bullet_0
= \delta$. From~\eqref{eqn:stat} we get $\pi^\bullet_k  =
\sum_{i=1}^nw_{ik}/\lambda_e$ for each $k=1,2\ldots,K$. From the
constraint (a), we have $\pi^\bullet_0+\sum_{i=1}^K \pi^\bullet_k = 1$ so that
$\sum_{i=1}^K \pi^\bullet_k = 1 -\delta$ under the active equality
constraint. Thus, we get $\lambda_e^\bullet =
\sum_{k=1}^K\sum_{i=1}^nw_{ik}/(1-\delta)$ from where we get
$\pi_k^\bullet = (1-\delta) \sum_{i=1}^nw_{ik}/
\sum_{k=1}^K\sum_{i=1}^nw_{ik} = (1-\delta)\pi_k^\ast/(1-\pi_0^\ast)$
for $k=1,2,\ldots,K$. The result follows upon combining the cases with
active and inactive inequality constraints. $\Box$

\subsubsection{Initialization and Classification}
\label{sec:init}
The {\it Rnd-EM}
approach~\citep{maitra09b} chooses -- from a pre-set number $M$ ($M=50$
in this paper) of ``valid'' 
randomly-realized initial values $\bTheta^{\ast}$ -- the configuration maximizing~\eqref{modobsllhd} and uses that
to initialize EM which is then run till convergence. We now discuss
the issue of choosing the initializing random realizations. We choose
each random initializer one component at a time. Specifically, we let
$(0.5, \bar\bv^\top)^\top$ be the initializing center for the first
(inactivated) component, where $\bar{\bv}$ represents the coordinates
of the central voxel in 
the image volume and the choice of 0.5 is from the mean of the uniform
distribution. For the initializing parameters of the beta
distributions of the activated components, we randomly select 
$K$ other values from among the $p$-values of voxels of potential
interest, for example, from voxels whose $p$-values were below some
$p_{\max}$ (say 0.05). 
The coordinates from each of these  $K$ voxels are our
initializers for these $K$ $\bmu_k$'s.

We now group each voxel in terms of the closest Euclidean 
distance of the observed $(p_i,\bv_i^\top)^\top$ to the potential initializing
candidates. 
(Note that the voxel coordinates $\bv$'s are scaled to lie in $[0, 1]$,
so that all voxel coordinates and $p$-values are of comparable
importance in the Euclidean distance for grouping. For each group $k$,
comprising the initially assigned $p$-values and voxel coordinates
$(p_{jk},\bv_{jk}^\top)^\top; j =1,2\ldots,n_k$, we set $\pi_k ={n_k}/n$ to be
the proportion of voxels assigned to the $k$th 
group and the remaining components of $\bTheta_k$ from the
maximizer of the log-density $\sum_{j=1}^{n_k}\log[
  b(p_{jk};\alpha_k,\beta_k)\phi(\bv_{jk};\bmu_k,\bSigma_k)]$ under
constraints (ii) and (iii).

We declare a candidate initializer ``invalid'' when a random initial 
grouping yields a single observation in some group, resulting in a
degenerate  initial $\bSigma_k$ for that group. (Invalid candidates
are discarded from further consideration.) We select the best 
initial value as 
$\bTheta^{(0)} = \argmax_{\bTheta^{\ast}}\ell(\bTheta^{\ast}; \bp, \bv)$ from
which the EM algorithm of the constrained mixture model is applied.
We consider a MLE valid if the EM algorithm of~\ref{sec:parameter_estimation}
converges to non-degenerate parameter estimates and leads to non-degenerate groups ({\em i.e.},each cluster has at least a few
observations, say $1 + c_v$).
The converged result from the best initial value does not
always lead to a valid MLE, so we  repeat the
{\it Rnd-EM} approach until we get a valid MLE. From the converged
valid MLE  $\hat{\bTheta}$, we estimate $\hat{w}_{ik}$ the {\em maximum a
posteriori} (MAP) evaluated at $\hat{\bTheta}$ and classify the $i$th voxel to
the $l$th component where $l = \argmax_k \hat{w}_{ik}$.

\subsubsection{Parallelizing EM for use with fMRI data}
\label{sec:APECM}

Let our APECMa  algorithm for maximizing \eqref{modobsllhd}  consist
of the $C=K+1$  cycles given by 
$
\mathbb{C} =
\{
\{\bpi, \bGamma_0\}, \allowbreak \{\bpi, \bGamma_1\}, \allowbreak \ldots,
\{\bpi, \bGamma_K\}\}
$
with each cycle containing a CM-step followed by a partial E-step (PE).
For example, PE-steps only update 
$u_{i,c}^{(s + [c|C])} \equiv
 f_c(p_i,\bv_i ;| \bGamma_c^{(s + [c|C])})$
for the $c$th cycle of the iteration $s$ where $c = 0,1,2,\ldots, C-1$.
The extra space here is
$((u_{ik}))_{1\leq i \leq n; 0\leq k \leq K}$ corresponding to the
$((w_{ik}))_{1\leq i \leq n; 0\leq k \leq K}$.
There is no need to update components other than the $c$th component because
$u_{ik}^{(s + [c|C])} \equiv u_{ik}^{(s + 1)}$ for $k < c$ have
been updated in a previous cycle of the same iteration $s$, while
$u_{ik}^{(s + [c|C])} = u_{ik}^{(s)}$ for $k > c$ have not changed
from the previous iterated value at $\bGamma_k^{(s + [k|C])}$.
Therefore,
\[
w_{ik}^{(s + [c|C])}
  =
  \frac{
    \pi_k^{(s + [c|C])} u_{ik}^{(s + [c|C])}
  }{
    \sum_{k = 0}^K \pi_k^{(s + [c|C])} u_{ik}^{(s + [c|C])}
  }
\]
for all $i$ and $k$, completing the E-step for the $c$th cycle of the $s$th
iteration.
The CM-steps are performed in turn on the $c$th cycle
and utilize $w_{ik}^{(s + [c|C])}$ in the same manner as the
M-step of the EM algorithm, but restricted to $\{\bpi, \bGamma_c\}$ for the
$c$th cycle.

The EGM algorithm requires no further changes to the PE-steps,
since we may distribute the dataset $\bX_{n\times (1+c_v)} \equiv
((p_i,\bv_i^\top)_{i,j})_{1\leq i \leq n; 1\leq j \leq (1+c_v)}$ in consecutive
row blocks
into $D$ cores for the EGM algorithm. For example, each core contains one
row block of the dataset,
say $\bX_{n_d \times (1+c_v)}$, with $n_d$ observations in the $d$th core and
$\sum_{d = 1}^D n_d = n$.
($n_d \approx n/D$ is ideally for gaining computation performance.)
Similarly, we distribute $w_{ik}$'s
and $u_{ik}$'s into the $D$ cores with dimensions $n_d \times (K + 1)$.
Therefore, the PE-steps in the data-distributed computed environment are the
same as in the serial environment and no communication across cores is
needed. Moreover, each core needs less computation -- about $1 / D$ on
the average -- for the EGM algorithm.

The CM-steps, however, need adjustment for application of the EGM algorithm.
For the mixing proportion, the unconstrained estimates are 
\[
\pi_{k\ast}^{(s + [c | C])}
  =
  \frac{
    \sum_{d = 1}^D \sum_{i = 1}^{n_d} w_{ik}^{(s + [c|C])}
  }{
    \sum_{j = 1}^K \sum_{d = 1}^D \sum_{i = 1}^{n_d} w_{ij}^{(s + [c|C])}
  }
\]
where
$
\sum_{i = 1}^{n_d} w_{ij}^{(s + [c|C])}
$
is the sufficient statistics calculated locally for each $d$ core
and is gathered/reduced in the G-step from all the other $(D- 1)$
cores. The constrained estimates for
$\pi_{k}^{(s + [c|C])}$ can be adjusted accordingly.
Unlike $\bGamma_k$'s, $\bpi$ is involved in every 
cycle so that all $\pi_k$'s need to be updated in each cycle
in order to fulfill conditions  that guarantee validity of the EM
steps, such as monotonicity of the
log likelihood and space-filling of the parameter space~\citep{Chen2013}.
For the beta distribution parameters with constraints, we numerically optimize
the objective function (in terms of $\alpha_c$ and $\beta_c$ for $c > 0$)
required by the \code{constrOptim()} function in \proglang{R}, such that
$
\sum_{i = 1}^{n} w_{i,c}^{(s + [c|C])}
\log b(p_i ; \alpha_c, \beta_c)
$
is maximized under the constraint (iii)
where
$
\sum_{i = 1}^{n_d} w_{i,c}^{(s + [c|C])}
\log b(p_i ; \alpha_c, \beta_c)
$
are the sufficient statistics of the objective function calculated
locally given each new numerical update of $(\alpha_c, \beta_c)$.
For the normal components, we compute
\[
\bmu_{c}^{(s + [c|C])}
  =
    \sum_{d = 1}^D \sum_{i = 1}^{n_d} w_{i,c}^{(s + [c|C])} \bv_i
   /
    \sum_{d = 1}^D \sum_{i = 1}^{n_d} w_{i,c}^{(s + [c|C])}
\]
and
$\bSigma_{c}^{(s + [c|C])}$ which is the diagonal matrix containing the diagonal
elements of
\[
\frac{
  \sum_{d = 1}^D \sum_{i = 1}^{n_d} w_{i,c}^{(s + [c|C])}
  (\bv_i - \bmu_c^{(s+[c|C])})(\bv_i - \bmu_c^{(s + [c|C])})^\top
}{
  \sum_{d = 1}^D \sum_{i = 1}^{n_d} w_{i,c}^{(s + [c|C])}
}
\]
where
$\sum_{i = 1}^{n_d} w_{i,c}^{(s + [c|C])} \bv_i$,
$\sum_{i = 1}^{n_d} w_{i,c}^{(s + [c|C])}$, and
$\sum_{i = 1}^{n_d} w_{i,c}^{(s + [c|C])} (\bv_i - \bmu_{c}^{(s + [c|C])})(\bv_i - \bmu_{c}^{(s + [c|C])})^\top$
are the sufficient statistics for the G-step.

\section{Performance  evaluations: Additional details}
\label{sec:simulations.supp}
\subsection{Illustrative examples to show the effect of the  identification complexity parameter}
\begin{figure*}
  \includegraphics[width=\textwidth]{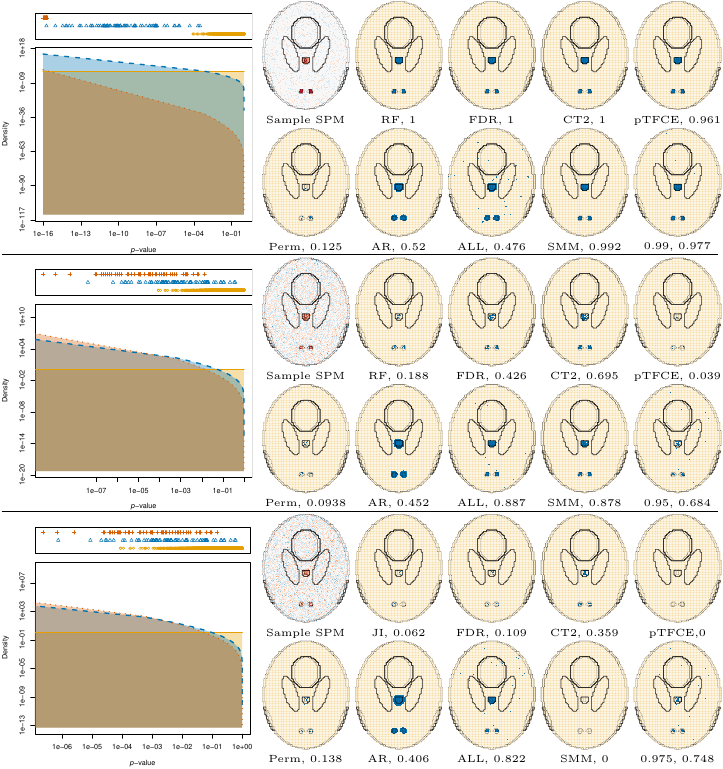}
  \caption{Sample realizations and activation detection for identification
    complexity $\omega=0.01$ (top set), $\omega = 0.50$ (middle set) and
    $\omega = 0.75$ (bottom set) on simulations obtained using the
    phantom in Figure~\ref{fig:phantom}a. For each set of values,
    we have, in the left  panel, a density of $p$-values,
    including a realization (placed atop 
    the density plots) of $p$-values using the generative model
    for $\pi_0=0.977$. The right panel displays a (i) sample
    SPM, obtained by calculating the upper quantile of the simulated
    $p$-values, as well as activation maps (along with their $\mJ$)
    obtained using  (ii) RF, (iii) FDR, (iv) CT-2nd, (v) pTFCE, (vi)
    permutation testing (Perm in the figure), (vii) AR-FAST, with
    $\alpha=0.01$, (viii) ALL-FAST, with $\alpha = 0.01$, (ix) SMM,
    and (x) MfnoX, with the best-performing $\delta$ from among
    \{0.95, 0.975, 0.99\}, and the choice indicated by the first value
    in the legend.} 
  \label{fig:simu_example_0}
\end{figure*}

\begin{figure*}
  \includegraphics[width=\textwidth]{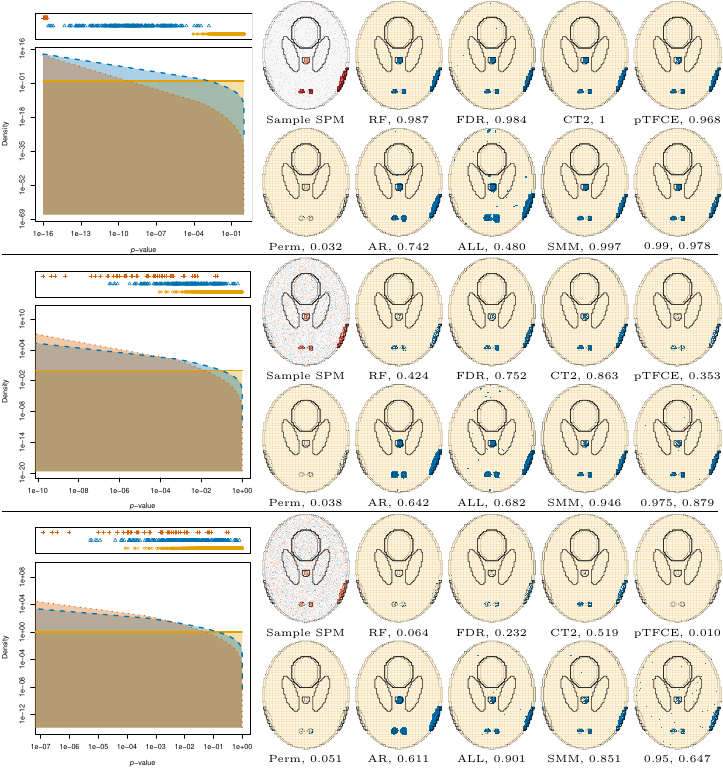}
  \caption{Sample realizations and activation detection for identification
    complexity $\omega=0.01$ (top set), $\omega = 0.50$ (middle set) and
    $\omega = 0.75$ (bottom set) on simulations obtained using the
    phantom in Figure~\ref{fig:phantom}b. For each set of values,
    we have, in the left  panel, a density of $p$-values,
    including a realization (placed atop 
    the density plots) of $p$-values using the generative model
    for $\pi_0=0.977$. The right panel displays a (i) sample
    SPM, obtained by calculating the upper quantile of the simulated
    $p$-values, as well as activation maps (along with their $\mJ$)
    obtained using  (ii) RF, (iii) FDR, (iv) CT-2nd, (v) pTFCE, (vi)
    permutation testing (Perm in the figure), (vii) AR-FAST, with
    $\alpha=0.01$, (viii) ALL-FAST, with $\alpha = 0.01$, (ix) SMM,
    and (x) MfnoX, with the best-performing $\delta$ from among
    \{0.95, 0.975, 0.99\}, and the choice indicated by the first value
    in the legend.} 
  \label{fig:simu_example_1}
\end{figure*}

\begin{figure*}
  \includegraphics[width=\textwidth]{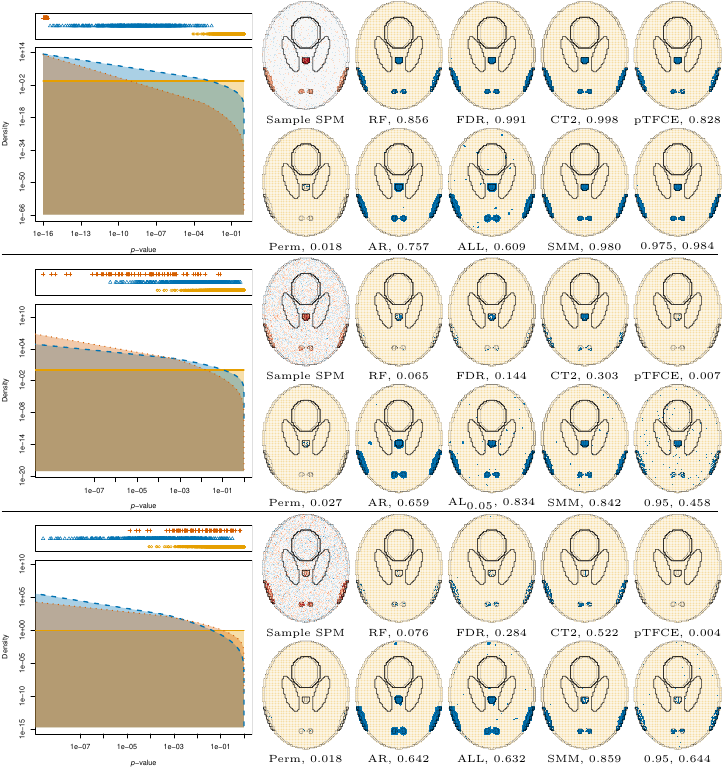}
  \caption{Sample realizations and activation detection for identification
    complexity $\omega=0.01$ (top set), $\omega = 0.50$ (middle set) and
    $\omega = 0.75$ (bottom set) on simulations obtained using the
    phantom in Figure~\ref{fig:phantom}c. For each set of values,
    we have, in the left  panel, a density of $p$-values,
    including a realization (placed atop 
    the density plots) of $p$-values using the generative model
    for $\pi_0=0.977$. The right panel displays a (i) sample
    SPM, obtained by calculating the upper quantile of the simulated
    $p$-values, as well as activation maps (along with their $\mJ$)
    obtained using  (ii) RF, (iii) FDR, (iv) CT-2nd, (v) pTFCE, (vi)
    permutation testing (Perm in the figure), (vii) AR-FAST, with
    $\alpha=0.01$, (viii) ALL-FAST, with $\alpha = 0.01$, (ix) SMM,
    and (x) MfnoX, with the best-performing $\delta$ from among
    \{0.95, 0.975, 0.99\}, and the choice indicated by the first value
    in the legend.} 
  \label{fig:simu_example_2}
\end{figure*}

We first provide sample illustrations (Figures~\ref{fig:simu_example_0},~\ref{fig:simu_example_1},
and ~\ref{fig:simu_example_2}) of the generative mechanism used
to simulate the $p$-values in our 2D simulation
The identification complexity parameter $\omega$ controls
the difficulty and the specificity of the activation problem and we
display sample densities, realizations and detected activation images
for representative  $\omega = 0.01, 0.5, 0.75$. Thus, the difficulty of our
activation detection  problem in our sample illustrations spans from
the moderately easy to the very difficult. As explained in Section~\ref{sec:2d}, the $p$-values are realizations from one of
$k$ densities given by  $\psi(p;\nu_k) =
\phi(\Phi_z^{-1}(1-p);\nu_k,1)/\phi_z(\Phi_z^{-1}(1-p))$. Note that
$\nu_0=0$ and at that value, the density reduces to the standard
uniform density (corresponding to the distribution of $p$-values from
the inactivated component). $\pi_0$, $\pi_1$ and $\pi_2$ are set to be
the proportion of pixels in each region for the given kind of
phantom. With these $\pi_0,\pi_1$ and $\pi_2$, and $\nu_1$ and $\nu_2$
are chosen to satisfy 
$\omega_{kl} = \omega$ for all $k\neq l \in \{0,1,2\}$, where
$\omega_{kl}$ is the overlap measure as defined in
\citet{maitraandmelnykov10}. The left panels of
Figures~\ref{fig:simu_example_0},~\ref{fig:simu_example_1},
and~\ref{fig:simu_example_2} show sample densities obtained for
$\omega = 0.01, 0.5, 0.75$, Simulated $p$-values are then realized from
$\psi(p;\nu_k)'$. Sample SPMs obtained from realizations of $p$-values
obtained from each of the realized densities in
Figures~\ref{fig:simu_example_0},~\ref{fig:simu_example_1} 
and~\ref{fig:simu_example_2} are also
displayed (in terms of values, atop each density figure) and
as the first image in the right panels.
\subsection{Illustrative example of performance in simulation settings}
\label{sec:simu_1}
The other images in the right panels of Figures~\ref{fig:simu_example_0},~\ref{fig:simu_example_1},
and~\ref{fig:simu_example_2} display, for each setting, and in order,
activation images obtained using our competitors (using values, where
appropriate, parameter values  that yielded the better
performance).
\bibliographystyle{IEEEtran}
\bibliography{b01-references,rm}

\end{document}